\documentclass[10pt,journal,compsoc, twoside]{IEEEtran}

\usepackage{graphicx} 
\usepackage{amsmath} 
\usepackage{booktabs}
\usepackage{subcaption}

\usepackage{xcolor} 

\title{On-line Application Autotuning Exploiting Ensemble Models}

\author{Tom\'{a}\v{s} Martinovi\v{c}, Davide Gadioli, Gianluca Palermo, 
Cristina Silvano
\IEEEcompsocitemizethanks{
\IEEEcompsocthanksitem T. Martinovi\v{c} is with IT4Innovations, V\v{S}B -- Technical University of Ostrava, Czech Republic.
\IEEEcompsocthanksitem D. Gadioli, G. Palermo, and C. Silvano are with Dipartimento di Elettronica, Informazione e Bioingegneria, Politecnico di Milano, Italy.
}
\thanks{This work is supported by the European Commission Horizon 2020 research and innovation program under grant agreement No 671623, FET-HPC ANTAREX and by the ESF in “Science without borders” project, reg. nr. CZ.02.2.69/0.0/0.0/16\_027/0008463 within the Operational Programme Research, Development and Education.}
}

\begin{document}
\IEEEtitleabstractindextext{%
\begin{abstract}
Application autotuning is a promising path investigated in literature to improve computation efficiency.
In this context, the end-users define high-level requirements and an autonomic manager is able to identify and seize optimization opportunities by leveraging trade-offs between extra-functional properties of interest, such as execution time, power consumption or quality of results.
The relationship between an application configuration and the extra-functional properties might depend on the underlying architecture, on the system workload and on features of the current input.
For these reasons, autotuning frameworks rely on application`s knowledge to drive the adaptation strategies.
The autotuning task is typically done offline because having it in production requires significant effort to reduce its overhead.

In this paper, we enhance a dynamic autotuning framework with a module for learning the application knowledge during the production phase, in a distributed fashion.
We leverage two strategies to limit the overhead introduced at the production phase.
On one hand, we use a scalable infrastructure capable of leveraging the parallelism of the underlying platform.
On the other hand, we use ensemble models to speed up the predictive capabilities, while iteratively gathering production data. 
Experimental results on synthetic applications and on a use case show how the proposed approach is able to learn the application knowledge, by exploring a small fraction of the design space.

\end{abstract}
}

\maketitle

\IEEEdisplaynontitleabstractindextext
\IEEEpeerreviewmaketitle


\section{Introduction}

With the end of Dennard scaling \cite{esmaeilzadeh2011dark}, the optimization focus shifted towards efficiency in a wide range of scenarios, from embedded to High-Performance Computing (HPC).
To this end, autotuning \cite{AutotuningInHPC} has been identified as a promising research.
In this direction, it is possible to use frameworks to optimize a specific task \cite{whaley1998automatically,puschel2004spiral} and frameworks to optimize knobs at either the system-level (e.g. the core frequencies) or the application-level (e.g. software parameters) \cite{ansel2014opentuner,rasch2017atf}.
Moreover, if the target application can tolerate an error on the output, \textit{approximate computing} \cite{mittal2016survey} represents an appealing path to further increase computation efficiency.
Finding application parameters that enable a quality-throughput tradeoff might be a complex task.
Therefore, several techniques have been proposed in literature to expose them, such as loop perforation \cite{hoffmann2009using} or task skipping \cite{rinard2006probabilistic}.

As a consequence of this trend, application requirements are increasing in complexity to address several extra-functional properties (EFPs), such as execution time, power consumption, and quality of the results.
On the other hand, application developers have started to expose in the source code a huge set of tunable parameters that alter the extra-functional behaviour of the application, thus enabling autotuning.
In this paper, we refer to these parameters as \textit{software-knobs}.

The relation between the software-knobs and the EFPs of interest for end-users is complex and unknown, therefore selecting the best configuration is a complex task.
The EFP values might depend on the underlying architecture, on the system workload, and on features of the current input.
A subset of software-knobs relates to parameters that aim at tailoring the application for the underlying architecture, such as work-group size, MPI runtime parameters or compiler options.
Typically, autotuning frameworks that address these parameters perform a Design Space Exploration (DSE) at design-time to find the most suitable configuration to be used in the production phase.
The main challenge in these approaches is  the exponential growth of the Design Space when considering several, usually unbounded, software-knobs. 
The second subset of software-knobs relates to application-specific parameters, such as the number of Monte Carlo trials, loop perforation factors or algorithm parameters.
Typically, it is easier to change these software-knobs during the production phase and their effects on EFPs are strongly coupled with the features of the current input.
For these reasons, autotuning frameworks that address these parameters typically model as an \textit{application-knowledge} the relationship among software-knobs, EFPs, and input features.
The autotuners leverage this knowledge to identify and seize optimization opportunities, improving the computation efficiency during the production phase.
The main challenge of these approaches consists of providing mechanisms to enhance the target application with an adaptation layer that gives self-optimization capabilities \cite{kephart2003vision}, considering application requirements and the system evolution.

Give the complexity of the tuning task, this is done typically offline prior to application execution, letting only the configuration selection at run-time. 
The main problem is that every time the code is ported to a new architecture, updated, or new input data are used for the elaboration, the offline tuning process should be redone.
However porting this phase online at production time requires a significant effort to minimize the tuning time and overhead as much as possible. 

In this paper, we propose a framework to learn online the application-knowledge at the beginning of the production phase. 
It has been designed to work in a distributed context were different entities can collaborate to the knowledge collection.
The framework mainly targets the context of HPC, where an application is composed of more than one process and it usually executes for a long period. 
However, it might be applied also in a wider range of scenarios. 
In fact, the benefits of learning the application-knowledge at runtime are the following: (1) we are able to leverage the parallelism of the platform to reduce the time-to-knowledge; (2) we are able to learn the behavior of the application using the actual input set and (3) using the actual execution environment.
Given that we are stealing time to the execution of the application, the main challenge addressed in this paper is to reduce as much as possible the time required to learn the application knowledge.
To reach this goal, in addition to the full exploitation of the parallel production machine, the proposed methodology.
We employ an iterative exploration strategy to reduce as much as possible the required number of samples.
In particular, we start to explore a fraction of the design space. If the derived models are not able to reach a target quality in the validation phase, the framework will resume the Design Space Exploration (DSE). Otherwise, the framework broadcasts the obtained application knowledge.

From the implementation point of view, we used the mARGOt \cite{gadioli2018margot} dynamic autotuning framework as starting point.
mARGOt is an adaptation layer that provides to the target application mechanisms to adapt in a reactive and proactive fashion, based on the application-knowledge.
In this paper, we enhance the mARGOt learning module with a model-driven approach.
The goal of this paper is not to compare different modelling techniques, but to leverage them for reducing the time required to compute the knowledge of the application.
We used synthetic applications with a known relation between EFPs and software-knobs to experimentally evaluate the out-of-sample predictions of the learning module. 
We focused on a molecular docking application, to assess the benefits of the proposed framework in a real-world case study.

The main contributions of the paper can be summarized as follows:
\begin{itemize}
\item We propose an autotuning framework to learn the relation between EFPs, software-knob configurations, and input features at the production phase;
\item We learn online how to exploit the parallelism of the production machine and to consider the production input data;
\item We leverage ensemble models to reduce the cost of the learning phase, while we select automatically the most suitable model according to the target problem; 
\item We enhanced a state-of-the-art autotuning framework with the proposed learning module. In particular, we extended it to leverage iterative exploration needed by the proposed approach;
\item We evaluated the framework with synthetic functions and a real-life case study from the HPC domain to demonstrate the introduced benefits. 
\end{itemize}

The remainder of the paper is structured as follows.
Section \ref{sec:related} compares the proposed framework with the related works, highlighting the main contributions.
Section \ref{sec:framework} describes the framework implementation, focusing on the relation among the involved components and showing the work-flow of the proposed methodology.
Section \ref{sec:formalization} formalizes the problem and describes in detail how the learning module computes the application knowledge.
Section \ref{sec:experimental} experimentally evaluates the proposed framework.
Finally, Section \ref{sec:conclusion} concludes the paper.

\section{State-of-the-art}
\label{sec:related}

In the context of autonomic computing  \cite{kephart2003vision}, an application is perceived as an autonomic element capable of self-management.
Among the self-* properties required by the self-management property, autotuning frameworks aim to provide self-optimization \cite{mahdavi2017systematic}.
In this context, the end-user specifies high-level requirements and the application should adapt accordingly, without the human-in-the-loop.
This is a promising path investigated in literature \cite{AutotuningInHPC}, where several autotuning frameworks have been evaluated according to their vision on how to provide self-optimization properties.
In the remainder of the section, we classify in six categories the most recent and relevant autotuning frameworks according to the methodology used to learn the application-knowledge.

In the context of HPC, there are several autotuning frameworks tailored to optimize specific domains.
ATLAS \cite{whaley1998automatically} was designed for matrix multiplication routine, FTTW \cite{frigo2005design} for FFTs operations, OSKI \cite{vuduc2005oski} for sparse matrix kernels, SPIRAL \cite{puschel2004spiral} for digital signal processing, CLTune \cite{nugteren2015cltune} and GLINDA \cite{Shen:2013:GFA:2482767.2482785} for OpenCL applications, Patus \cite{christen2011patus} and Sepya \cite{kamil2012productive} for stencil computations.
Considering domain-specific autotuning, these works are interesting.
However, they were designed to take orthogonal decisions with respect to the proposed approach, which is oriented towards supporting the online-learning phase.

The second category represents frameworks that aim to apply code or binary transformations to introduce the possibility of exploiting accuracy-throughput tradeoffs.
QuickStep \cite{misailovic2013parallelizing}, Paraprox \cite{samadi2014paraprox}, and PowerGAUGE \cite{dorn2017automatically} are some examples in this category.
The main focus of these works is on how to expose tradeoffs by introducing software-knobs. 
The parameter tuning phase is done at design-time by relying on the representative input set.

The third category includes frameworks that aim to explore a very large design space, to find the best configuration of the software-knobs according to the application requirements before the production phase.
ATune-IL \cite{schaefer2009atune}, OpenTuner \cite{ansel2014opentuner}, and the ATF framework \cite{rasch2017atf} are some examples in this category.
ATune-IL provides a mechanism to prune and reduce the configuration space according to the code structure and to the dependencies among software-knobs.
OpenTuner uses a multi-armed bandit framework to choose the best search algorithm for the given application.
ATF framework improves the OpenTuner strategies by considering also domain-constraints of the parameters.
Given that the tuning phase is done at design-time, these frameworks usually target software-knobs loosely-coupled with the inputs.
Moreover, the output of the tuning process is a single software-knob configuration, not the application-knowledge required to adapt at the production phase.

In the fourth category, we represent frameworks that target streaming applications.
They typically learn the application-knowledge at design-time, to be leveraged during the production phase.
The Green framework \cite{baek2010green}, the Sage \cite{samadi2013sage} framework, and PowerDial \cite{hoffmann2011dynamic} are some examples in this category.
The main focus of these works is on how to provide reaction mechanisms to a streaming application.
Therefore, they use representative inputs to derive the application knowledge and then react during the production phase.
This approach assumes only a few abrupt changes in the input that must be elaborated.
However, this assumption might not hold in all the types of workloads, e.g. see Section \ref{ssec:geodock} or \cite{ding2015autotuning, sui2016proactive, gadioli2018margot}.

In the fifth category, we consider the autotuning frameworks that adapt an application also in a proactive way by using input features and learning the application-knowledge at design-time.
Petabricks \cite{ding2015autotuning}, the framework proposed in \cite{guo2003bayesian}, and Capri \cite{sui2016proactive}, are some examples in this category.
The methodology used to derive the application-knowledge assumes the possibility to select which input to consider in the representative set used during the learning process.
The methodology proposed in this paper has been designed to use directly the production input, therefore it is not possible to apply the same approaches.
Moreover, these frameworks are able to express a tradeoff between a quality metric and an additional EFP only.
On the other hand, the proposed approach is capable to address an arbitrary number of EFPs.

In the sixth and last category, we represent the autotuning frameworks (such as  \cite{laurenzano2016input, miguel2016anytime}) that adapt an application also in a proactive way, without learning the application-knowledge at design-time.
The framework proposed in this paper falls in this category.
The IRA framework \cite{laurenzano2016input} defines the concept of \emph{canary input} as the smallest sub-sampling of the actual input, which has the same property as the original input. 
It proposes the usage of a canary input for a runtime parameter exploration of the target application for each data to be processed. 
Then, it uses the fastest configuration of the software-knobs resulting within a given bound on the minimum accuracy.
The main drawback of this methodology is that the presented sub-sampling technique applies to matrix-like input, therefore limiting the applicability of the framework.
A rather different approach with respect to the previous ones is Anytime Automaton \cite{miguel2016anytime}.
It suggests source code transformations to re-write the application by using a pipeline design pattern.
The idea is that the longer the given input executes in the pipeline, the more accurate the output becomes.
The work targets hard constraints on the execution time, therefore the idea is to interrupt the algorithm when it depletes the time budget.
In this way, it is possible to have guarantees on the feasible maximum accuracy. However, this approach works only for exploiting accuracy-performance tradeoffs.

In this paper, we propose a framework that can be considered orthogonal to most of the autotuning frameworks presented so far. In particular, it has been designed to support dynamic tuning at runtime with the capability to learn online the application-knowledge. To reduce the learning time, it uses two different approaches. First, it is based on a scalable infrastructure capable of exploiting the parallelism of the underlying production platform. Second, it uses ensemble models for accelerating the learning phase, while selecting always the best model for the target problem.

\section{Proposed Framework Architecture}
\label{sec:framework}

\begin{figure}
    \centering
    \includegraphics[height=4.8cm]{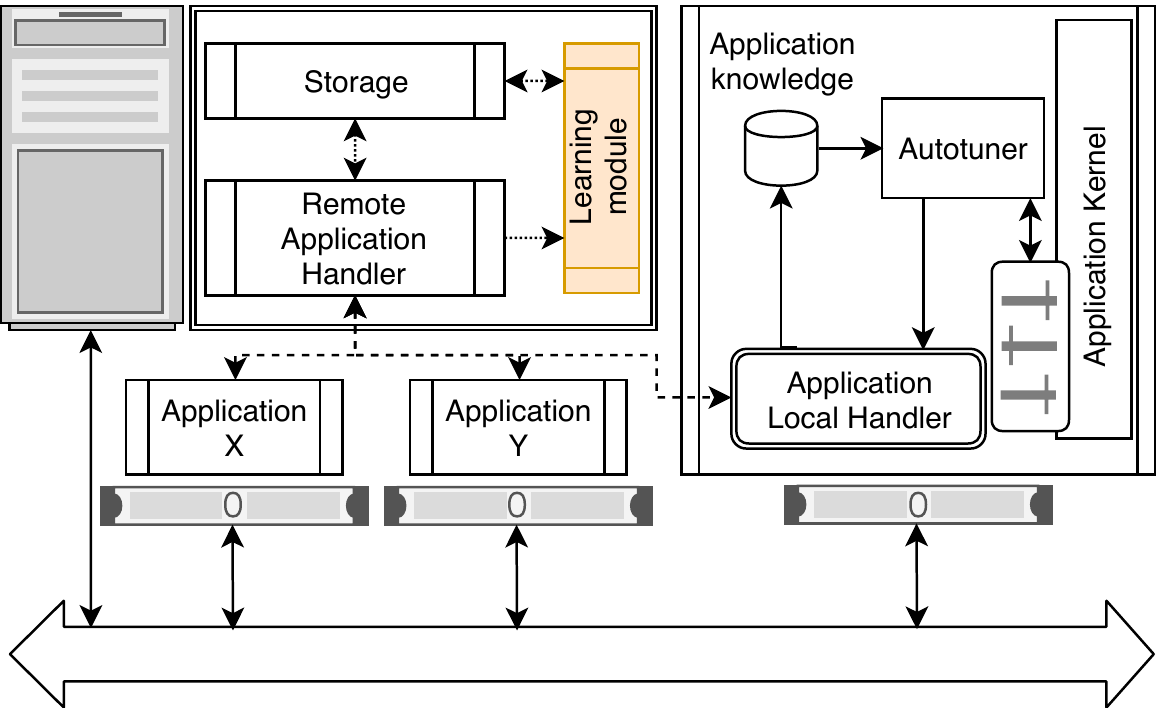}
    \caption{Overview of the proposed autotuning framework. Model analysis and out-of-sample predictions are computed by the learning module, executing in a dedicated server.}
    \label{fig:agora}
\end{figure}

This Section introduces the architectural view of the proposed technique in the context of the whole adaptive framework. Later, Section \ref{sec:formalization} will focus on the learning approach from the methodology point of view.

The proposed approach uses as a starting point the mARGOt autotuning framework \cite{gadioli2018margot}, which aims to enhance a target application with an adaptation layer to provides mechanisms to adapt in a proactive and reactive fashion.
From the implementation point of view, mARGOt is a C++ library to be linked to a target application.
Therefore, each instance of the application can take autonomous decisions.
mARGOt takes as input an application-knowledge defined as a discrete set of \textit{Operating Points} (OPs).
Each OP relates a software-knobs configuration with the expected EFP values reached using the configuration, according to a set of input features.
Therefore an OP is composed of three sets of values, namely  $<\overline{K},\overline{M},\overline{F}>$: the software-knobs configuration ($\overline{K}$), the expected EFPs ($\overline{M}$), and the related input feature values ($\overline{F}$).
By using this representation, the mARGOt autotuner framework is capable to select the most suitable one, according to application requirements defined as a constrained multi-objective optimization problem.
Moreover, it uses feedback information from monitors to adapt in a reactive way, while it uses features of the current input to adapt in a proactive way.

The main goal of the proposed framework is to obtain the OPs list during the production phase, without requiring a design-time profiling phase.
The proposed methodology guides the learning process by leveraging the underlying mARGOt infrastructure.
To achieve these goals, we exploit the possibility to dynamically update the OPs list of each application client.
In particular, we would like to assign at each application client a different software-knob configuration to distribute the design space exploration.
We would like to broadcast the OPs list to all the application clients, once the learning module generates the application knowledge.
In this way, 1) it is possible to leverage all the available nodes to reduce the time-to-model, 2) the application knowledge is tailored for the current input, and 3) we measure the EFPs with the production environment.

Figure \ref{fig:agora} provides an overview of the proposed autotuning framework.
It is composed of three components: a \textit{remote application handler} (server), an \textit{application local handler} (client) and the \textit{learning module}.
The \textit{remote application handler} is the central coordinator, implemented as a thread pool that executes in a dedicated server.
To store information about the managed applications and the status of the DSE, the server might use the Apache Cassandra database for scalability reasons, or CSV files for small instances.
The \textit{learning module} is the core of the proposed approach and it performs three main tasks:
1) It leverages the Design of Experiments techniques to sample efficiently the design space to be explored;
2) It leverages state-of-the-art modelling techniques to interpolate out-of-sample predictions;
3) It uses a validation stage to test whether the quality of the obtained model is acceptable or it requires additional effort to explore the design space. An extensive description of the learning module is presented in Section \ref{sec:formalization}.
The \textit{application local handler} is a service thread in each application instance which runs asynchronously with respect to the application execution flow. 
Its main goal is to manipulate the application-knowledge of the related application instance.
In particular, during the design space exploration phase, the \textit{application local handler} forces the autotuner to select the software-knobs configuration that must be evaluated.
When the model is available, it sets the application-knowledge accordingly.
Moreover, it has in charge the synchronization with the server counterpart.
The communications between server and clients leverage the MQTT or MQTTs protocols.

\begin{figure}[t]
\centering
    \includegraphics[width=0.85\columnwidth]{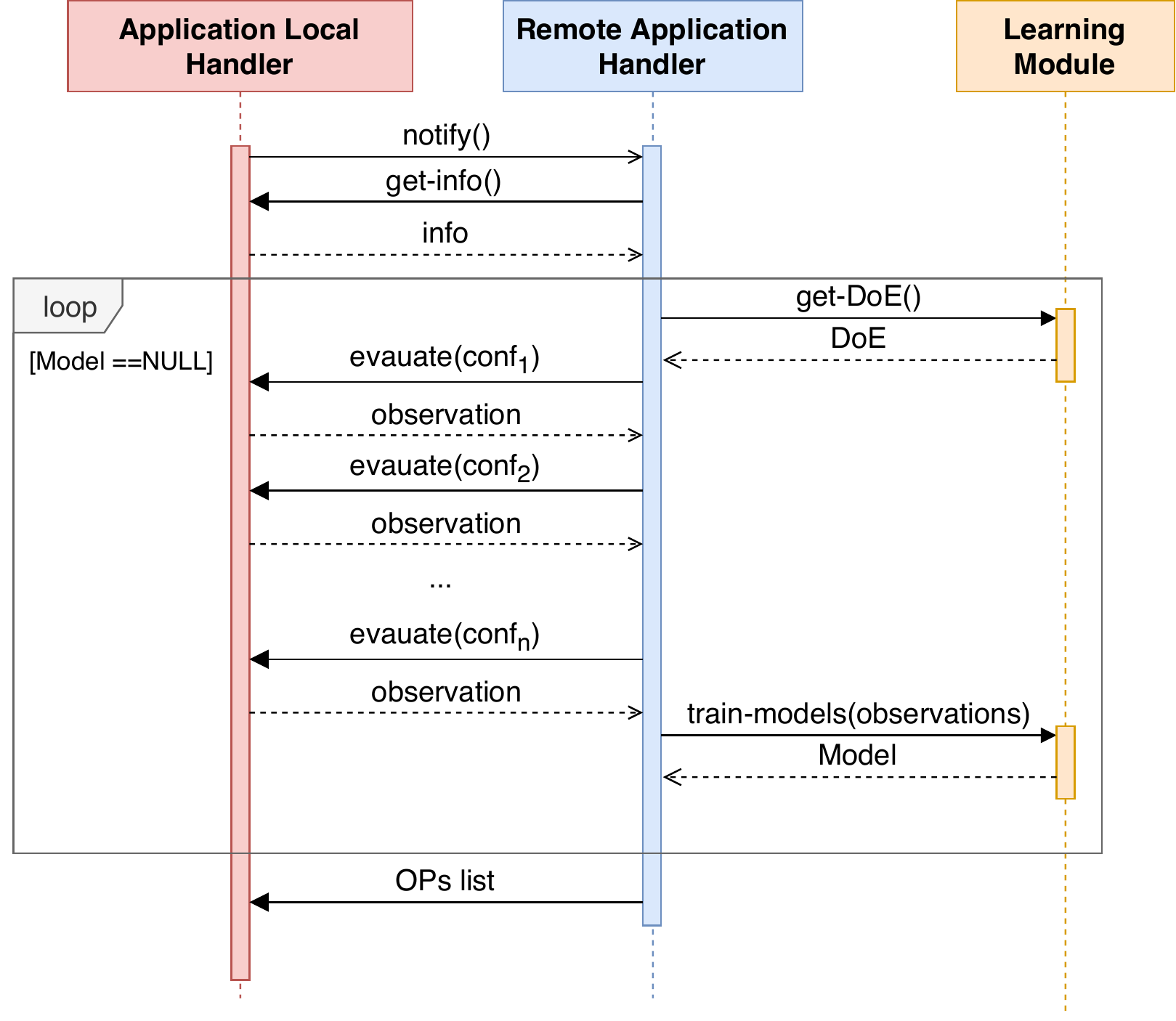}
    \caption{Sequence diagram of the interaction between the remote application handler, the application local handler, and the learning module. In this diagram, we consider an unknown application composed of a single process.}
    \label{fig:uml_procedure}
\end{figure}

Figure \ref{fig:uml_procedure} shows the typical workflow of the framework when it interacts with an unknown application.
In particular, it is composed of the following steps:
\begin{enumerate}
    \item The clients notify themselves to the server.
    \item The server asks one client information about the application, such as the number of software-knobs and their domain.
    \item \label{step3} Once the server has collected the information, it will call the learning module to generate a set of configurations to explore (DoE, design of experiments).
    \item The server dispatches to the available clients the configurations to evaluate in a round-robin way.
    \item Once the clients have explored all the configurations, the server requests a model from the learning module.
    \item The learning module trains, validates, selects and returns the best model. If the model is not valid, it returns an empty model.
    \item If there is a valid model, the server broadcasts the model to the available clients, otherwise it restarts from step \ref{step3}.
\end{enumerate}

The framework implementation is resilient to crashes at the server-side and at the client-sides.
Moreover, whenever a new client becomes available, it can join the design space exploration or receive the model directly.

\section{Proposed Methodology}
\label{sec:formalization}

\begin{figure*}[ht]
\centering
    \includegraphics[width=\textwidth]{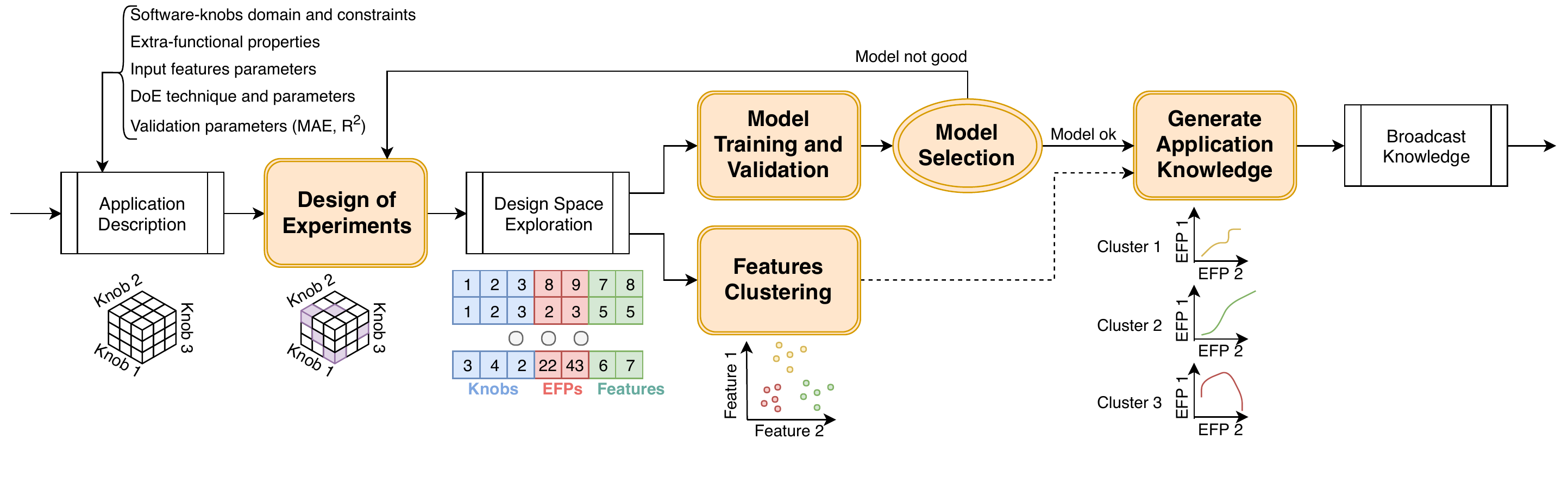}
    \caption{Overview of the proposed methodology: Orange elements represent components of the \textit{learning module}, white elements represent components of the mARGOt autotuning framework.}
    \label{fig:flow}
\end{figure*}

The goal of the proposed approach is to learn, during the production phase, the relation between software-knobs configuration, EFPs, and input features.
The main challenges in achieving this goal are twofold.
On one hand, we need to reduce as much as possible the time for learning the application-knowledge.
On the other hand, we are not capable to control the features of the input.
While we are capable to force an application instance to use a given software-knobs configuration, the input set is the one of the production run.

Figure \ref{fig:flow} shows an overview of the proposed methodology.
Orange elements represent components of the \textit{learning module} that drives the learning process, white elements represent components of the mARGOt framework.
Given the description of an unknown application, we use techniques of Design of Experiments (DoE) to sample efficiently the design space.
After the exploration of the selected software-knobs configuration, we do two operations.
First, we build a model for each EFP of interest.
Second, we cluster the observed input features.
If the best model that we found is deemed valid, we use it to generate the list of OPs to broadcast to the application clients.
Otherwise, we generate additional software-knob configurations to be evaluated and we restart the procedure.
The remainder of this section formalizes the main components of the proposed approach: the DoE techniques, the modelling techniques, the model training and validation, the model selection and finally the input feature clustering.

\subsection{Design of Experiments}
\label{ss:DOE}

The proposed approach aims at obtaining the application knowledge at the production phase, therefore we want to reduce the design space exploration as much as possible.
To reach this goal, it is important to sample the design space to maximize the retrieved information.
This is a well-known problem in literature, where several design of experiments (DoE) techniques were investigated \cite{montgomery2017design}.
The proposed framework leverages the Dmax algorithm \cite{Dupuy2015}, which maximizes the determinant of the correlation $\rho_{ij}$ defined as in Equation \ref{eq:dmax}:
\begin{equation}
\label{eq:dmax}
    \rho_{ij} = \begin{array}{cc}
         1 - \gamma &  \mathrm{if}~h_{ij} \leq \varepsilon,\\
         0 &  \mathrm{if}~h_{ij} > \varepsilon, 
    \end{array}
\end{equation}
where $h$ is the distance between points $x^i$ and $x^j$, $\varepsilon$ is the threshold distance of the correlation between two points and $\gamma$ is a variogram.
This measure maximizes the information entropy of the design and thus strives to maximize the information gained by exploring the given software-knobs.
Therefore, this method works well for creating DoE on the design space with no apriori knowledge.

This DoE technique exposes two free parameters: the total number of software-knobs configuration to be explored $n$ and the threshold distance $\varepsilon$.
We provide to the end-user the possibility to change these parameters from their default values ($40$ and $0.2$ respectively).
Moreover, the end-user might specify how many times each selected software-knob configuration is explored.
Ability to make several executions of the same software-knob is important to increase the robustness of the estimation in cases of the non-deterministic applications.
Given that the input features are not controllable, multiple runs of the same software-knobs configuration might lead to learning the knowledge from different feature sets.

The Dmax algorithm is designed to sample a continuous design space.
The application description defines a discrete domain for each software-knob, therefore the selected samples are then mapped to the closest non-selected software-knobs configuration.
Moreover, to accommodate the needs for more complicated design spaces, it is possible to set restrictions on the software-knobs domain.
There are several ways to implement restrictions on a design space.
On one hand, it is possible to compute the feasible design space based on the provided restrictions.
However, in the case of nonlinear relationships between software-knobs, this would lead to hard-to-solve nonlinear inequalities.
On the other hand, it is possible first to create the full factorial design space, and then to remove software-knobs configuration that does not satisfy the given restrictions.
The latter approach is feasible since the user-defined design space is discrete and finite.
In the proposed approach, we select the software-knobs configurations to explore in four steps:
\begin{enumerate}
    \item Create a full-factorial design.
    \item Remove the OPs not valid for the given restrictions.
    \item Use Dmax to sample the continuous design space to explore.
    \item Map the selected samples to the software-knobs domain.
\end{enumerate}

The output of this stage is a list of software-knob configurations to explore in the Design Space Exploration.
The latter task is performed by the autotuning framework.

\subsection{Modelling Techniques}
\label{ss:models}

This section describes the model families used to learn the relation between EFPs, software-knobs, and input features.
The learning module models each EFP independently.
Therefore, in our notation $\hat{y}$ represents the expected value of the target EFP, while $x$ represents the vector of predictors, i.e. software-knobs and input features.
In particular, we model an EFP as $\hat{y} = g(x)$, where the function $g$ is represented by a given modelling technique.
The remainder of the section describes the type of models used in the proposed approach.

\subsubsection{Linear Models}

Linear regression \cite{montgomery2017design} with $n$ dependent variables and $p$ explanatory variables is defined in Equation \ref{eq:linear}:
\begin{equation}
\label{eq:linear}
    \hat{y} = \alpha  X \beta + \varepsilon,
\end{equation}
where $\alpha$ is a constant, $\beta$ is a vector of $n$ parameters, $X$ is a $n \times p$ matrix of explanatory terms, and  $\varepsilon$ is vector of residuals or errors.

We use two types of linear models: first we consider only the model with a constant and the explanatory variables (\emph{first order}); second we consider also two-way interactions of explanatory variables (\emph{first order with interactions}).

\subsubsection{MARS Models}

The second family of models used in the learning module is multivariate adaptive regression splines (MARS) \cite{Friedman1991}.
This model iteratively adds basis functions to create the best possible representation of the variables interactions (non-parametric model).
The MARS representation is defined in Equation \ref{eq:mars}:
\begin{equation}
    \hat{y} = c + \sum_{i=1}^k w_i B_i(x),
    \label{eq:mars}
\end{equation}
where $c$ is a constant, $k$ is number of basis functions, $w_i$ is the constant coefficient of the basis function $i$, $B_i(x)$ is the basis function $i$.
The basis function is of the form $\max(0, d_i-x)$, $\max(0, x-d_i)$ or the multiplication of multiple basis functions.
The parameter $d_i$ is a constant estimated by the model.
In the learning module, we also use a variation of this model, named POLYMARS, which enables the maximum of two-way interactions in the model \cite{Stone1997}.

\subsubsection{Kriging Model}
The third family of models used by the learning module is Kriging \cite{kriging}.
We use an extension of the original model, named Universal Kriging (UK) \cite{WEBSTER1980}, which assumes that observed values $y$ come from a deterministic process $Y$ given by Equation \ref{eq:krigin}:
\begin{equation}
    Y(x) = \mu(x) + Z(x)
    \label{eq:krigin}
\end{equation}
where $\mu$ is a trend defined by the number of basis functions and $Z$ is a known covariance kernel.

In the context of the proposed framework, the generating process is seldom deterministic, therefore we need to relax this assumption.
In particular, we forced the determinism by averaging the observed values for each observed software-knobs configuration.

\subsubsection{Ensemble Models}
\label{ss:ensembles}

Model ensembling is a well-known approach to increase the predictive capabilities of base models by combining them together, using different techniques.
The learning module leverages two techniques based on cross-validation models: bagging \cite{Breiman1996_bagging} and stacking \cite{Breiman1996}.

\begin{figure}[t]
\centering
    \includegraphics[width=0.8\columnwidth]{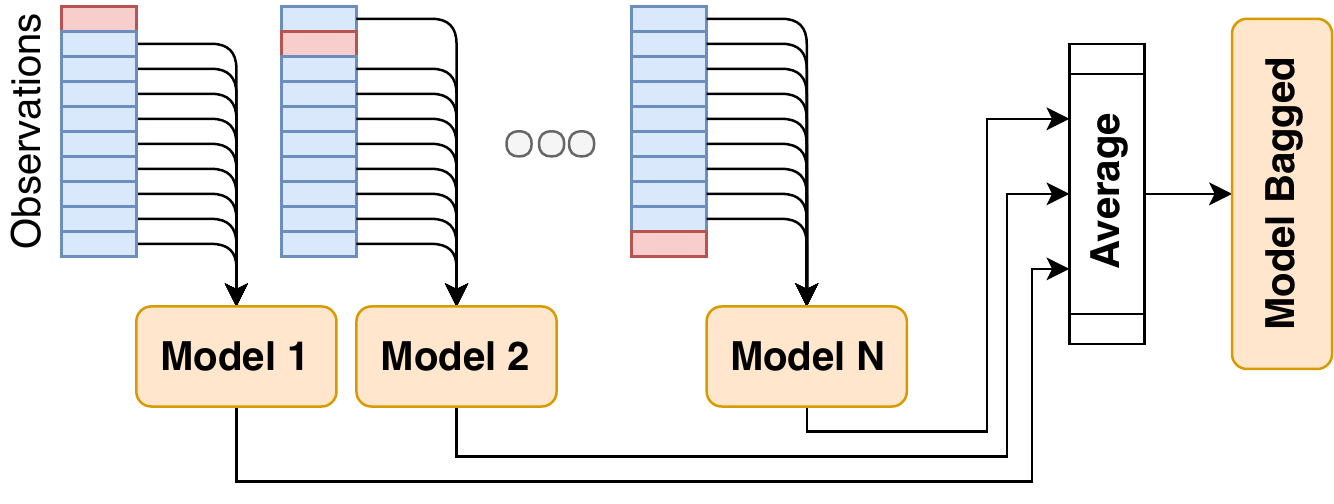}
    \caption{The procedure to compute ensemble models by using the bagging approach, starting from $N$ observations.}
    \label{fig:bagging}
\end{figure}

The bagging approach aims at decreasing the variance of the prediction.
It focuses on a single base modelling technique and it combines instances of the model trained with different data sub-samples.
The basic idea of bagging is increasing the robustness of the predictive capabilities of the model by combining models trained on the datasets with only small differences.
For example, if we have 10 observations and train 10 different models always leaving out one observation, each model pair will share 8 out of 9 observations used for training the model.
Combination of such models should lead to a more robust ensemble model.
To obtain the prediction, it uses the mean of the predictions generated by the model instances.
Figure \ref{fig:bagging} shows the process of bagging model computation.

The stacking approach aims to increase the robustness of the prediction by combining together base models.
A stacked model should be able to decrease the weaknesses of the individual models and leverage their strengths.
The learning module uses a weighted mean of the model families describes in the previous section together.
The weights for each model family aims at minimizing the error of the stacking model towards the training data observations.
Moreover, they must be positive and sum up to one.
This is a quadratic optimization problem and it has been solved using a dedicated R package \cite{quadprog}.
\begin{figure}[t]
\centering
    \includegraphics[width=\columnwidth]{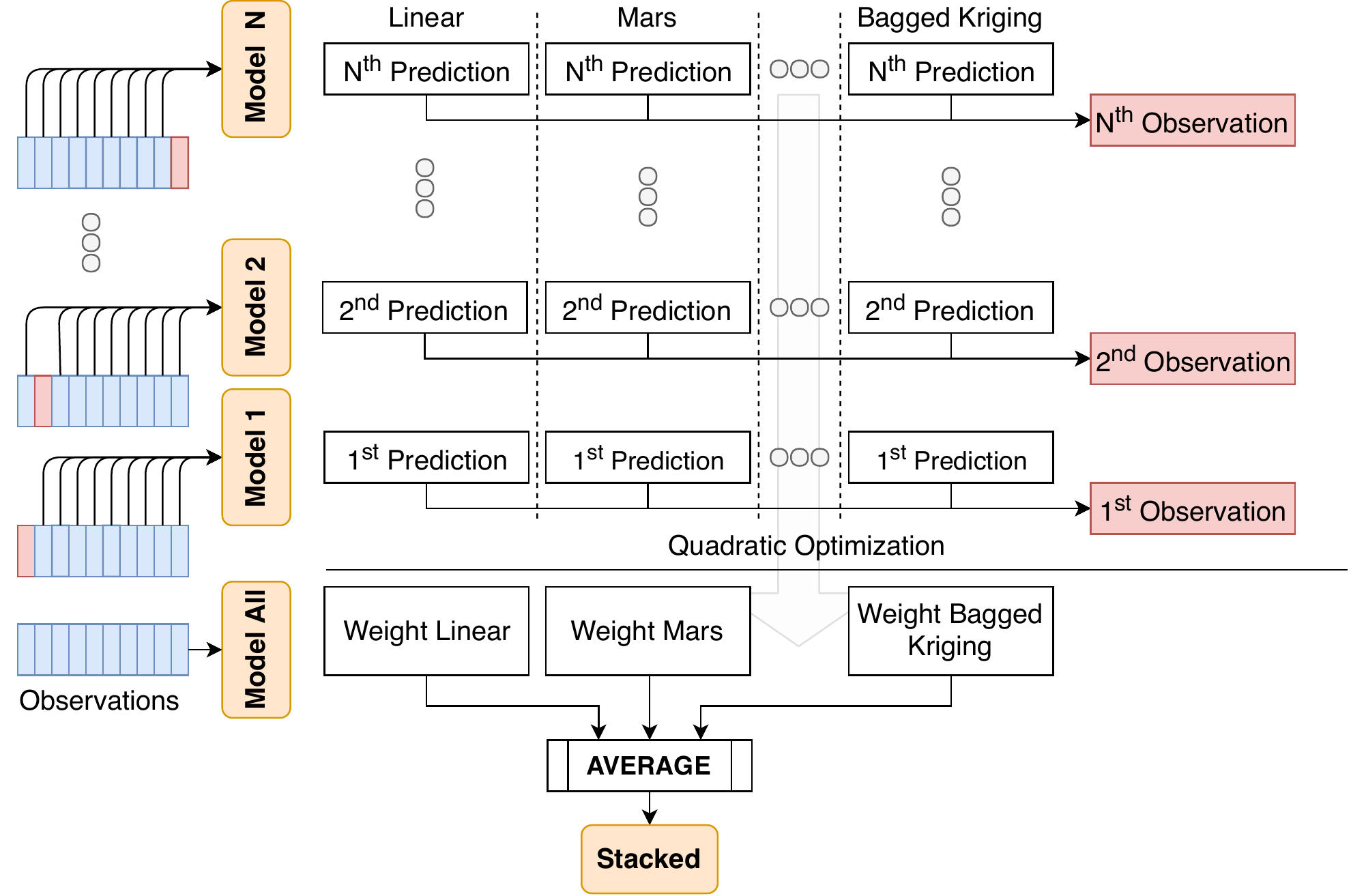}
    \caption{The procedure to compute ensemble models by using the stacking approach, starting from $N$ observations.}
    \label{fig:stacking}
\end{figure}
Figure \ref{fig:stacking} shows the process used by the learning model for computing the stacking model, based on the following steps:
\begin{enumerate}
    \item Train several instances of each model family using a cross-validation scheme.
    \item \label{step:stacking_predictions}Use every instance of a model family to predict the holdout data for the training.
    \item \label{step:stacking_matrix}Create a matrix of predicted data considering the predictions of step \ref{step:stacking_predictions} for all the model types.
    \item \label{step:compute_weights_stacking}Compute the weights for each model family by using quadratic optimization on the matrix of prediction computed at step \ref{step:stacking_matrix}, to best fit the training data.
    \item \label{step:total_training}For each model family, train the model by using the complete training data.
    \item Average the prediction of models trained in step \ref{step:total_training} using the weights computed in step \ref{step:compute_weights_stacking}.
\end{enumerate}

As stated in the previous work \cite{Breiman1996}, this definition of the stacking greatly reduces the exploration space and makes the weights estimation robust.

\subsection{Model Training and Validation}
\label{ss:validation}

This section describes how we partition the information from the Design Space Exploration to train and to validate the models.
This step is crucial to broadcast to the application clients a reliable application knowledge.

The typical approach for testing how models fare in the prediction is to divide the input data in a training set and validation set.
However, given that the proposed framework leverages the application-knowledge during the production phase, we might have a small set of observation for training and for validating the models.
This is true especially for the first iterations of the learning process. 
Therefore, the learning module uses three different validation schemes according to the number of software-knob configuration explored and a parameter $v_f$ that represents the validation set ratio.
The selected approach depends on the relationship of $v_f$ and $(n-k)/n$, where $n$ is the number of explored software-knob configurations and $k$ is the number of cross-validation folds.

In the general case, when $ (n-k)/n \geq v_f$, we use a k-fold validation scheme: the full set observations are divided into $k$-parts of equal size.
One part is always used as a holdout set and the rest is used to train the model.
This implies that the learning module trains $k$ models and each of them will have out-of-sample predictions on a different part of data.
We will call these models \textit{cross-validation models}.
In this case the validation set has size of $n \times v_f$.

On the other cases, when there are only few explored configurations (i.e. $ (n-k)/n < v_f$), then we use the first $k$ data to train the \textit{cross-validation models} using a leave-one-out cross-validation scheme.
We use the remainder of the data as the holdout validation set.
Moreover, If the number of explored software-knob configurations is less than $k$, we apply a leave-one-out cross-validation, without using any holdout validation set.
In this case, we are not using ensemble models due to bias problems in the model selection.
On one hand, bagged models use the average of \textit{cross-validation models}.
On the other hand, stacking model requires a training with all the data used to compute \textit{cross-validation models}.
Therefore, without any holdout data, ensemble models are implicitly trained by using the full set of observations available.

Using a holdout validation set gives better information about how the models fare on the out-of-sample prediction.
Moreover, it allows us to use ensemble models which are based on the cross-validation models.
In practice, the third method is implemented only for the special cases when the user decides to use really high $k$ or he has just a few observations.
It is important to notice how in this case the results of the models on out-of-sample predictions might be highly volatile.

To increase the robustness of the model's prediction capabilities estimation, we used a similar approach to the k-fold cross validation also with the holdout validation set.
We split the input data into $l = \mathrm{floor}(n/m)$ folds, where $m$ is a number of observations used for the holdout validation set.
Then validation is made for each of the $l$ folds.
In this way, it is possible to test prediction capabilities on the whole input dataset and not just one randomly selected part for the validation.

\subsection{Model Selection}
\label{ssec:model_validation_selection}

To quantify the prediction quality of a model, we consider two metrics.
A variant of the coefficient of determination ($R^2$)  \cite{Everitt2010}, and the mean absolute error, normalized by the observed values range ($MAE\_adj$).
In our case, $R^2$ is computed as the square of the correlation between observed and predicted data.
This variant \cite{Everitt2010} has been selected because it is restricted to [-1; 1] and it can be used on the cross-validation and out-of-sample predictions to compare the results in a consistent manner.
For evaluating these metrics for each base modelling types, we consider the mean of $R^2$ and $MAE\_adj$ across the \textit{holdout validation models}.
In the special case when a number of explored software-knobs is less than $k$ and the holdout set is not used the mean of $R^2$ and $MAE\_adj$ of the \textit{cross-validation models} is used. 
For model ensembles, we compute them considering the whole set of observations.

Once we evaluate all the models, we deem as eligible the ones that have  $R^2$ higher than $\epsilon_r$ and $MAE\_adj$ less than $\epsilon_m$, to enforce a minimum quality.
Among the eligible models, we select the one that minimizes the $MAE\_adj$.
If no model is eligible, the proposed approach will restart the design space exploration, up to a maximum number of iterations.
When the maximum number of iterations ($maxIt$) is reached, the learning module selects the model with the smallest $MAE\_adj$ for the out-of-sample predictions.
The parameters $\epsilon_r$, $\epsilon_m$ and $maxIt$ are exposed to end-user and by default, they are set to $0.5$, $0.1$ and $-1$ respectively.

\subsection{Feature Clustering}
\label{ss:clustering}
The main goal of this component is to find representative clusters based on the input features to be exploited in the application-knowledge.
This stage is based on the same considerations done in Petabricks \cite{ding2015autotuning}, that inspired our implementation.
Algorithmic choices and software-knob configurations are often sensitive to input features. However, the feature space can be too large to be completely considered. 
Clustering techniques are an intuitive solution to reduce the problem complexity.

To cluster the input features observed in DSE, we apply well-known clustering techniques such as k-means \cite{hartigan1979algorithm} and DBSCAN \cite{ester1996density} according to a parameter exposed to the end-user.
K-means is a clustering technique to minimize the intra-cluster variance, where the user sets a fixed number of clusters.
Using a different approach, DBSCAN partitions the samples in clusters, according to their proximity, where the user sets a fixed distance threshold.
Moreover, the user is able to manually define the clusters.

This component is activated after the Design Space Exploration.
As shown in Figure \ref{fig:flow}, it works in parallel with the model learning component.
Once the best model has been selected, the generated clusters on the input features are combined with the learned models and used to generate the application-knowledge to be broadcast to the clients.

\section{Experimental Results}
\label{sec:experimental}

This section describes the experimental assessment of the proposed framework.
First, we evaluate the prediction capabilities of the proposed framework, by using synthetic applications with a known relation between the EFPs and the software-knobs.
Then, we focus on a geometrical docking application to evaluate the benefits provided by the proposed framework for the end-user.
For this experiment, we used a platform with eight CPUs Intel(R) Xeon(R) X5482 @3.20GHz and 8GB of memory.
Concerning the implementation, we used the R packages for the Dmax design \cite{Dupuy2015} and for estimating the linear models \cite{linear}, the MARS \cite{MARS}, POLYMARS \cite{POLYMARS}, and Kriging \cite{Roustant2012} models.
The plugin uses the tidyverse package \cite{tidyverse} for data manipulation and to unify the workflow of different models.

\subsection{Framework Validation with Synthetic Applications}
\label{ssec:validation}

The main goal of this section is to evaluate the proposed framework in out-of-sample predictions, when trained with a fraction of the design space.
This is the main challenge of the proposed framework, given that each sample used to train the model steals time from the target application.
To validate the proposed framework, we created two synthetic applications with a known relation between the software-knobs and the EFPs that has been inherited from well-known functions \cite{Binh1999, Kursawe1991}. 

The first application is derived from the work of Binh \cite{Binh1999}, and it has been defined in Eq. \ref{eq:binh}:
\begin{equation}
\label{eq:binh}
\begin{array}{lrl}
   b_1 (x,y) &=& x^2 - y \\[3pt]
   b_2 (x,y) &=& -0.5x - y - 1\\[3pt]
   \mathrm{where} && -7 \leq x,y \leq 4.
\end{array}
\end{equation}
where $b_1(x,y)$ and $b_2(x,y)$ represent the EFPs of interest for the end-user, while $x$ and $y$ represent the software-knobs of the application.
The mathematical formulas describing the two functions represent the relationship between the EFPs and the software-knobs modeled by the framework by using a limited training set.
The proposed approach models each EFP independently, therefore this first application aims to demonstrate the ability of the framework to model linear and nonlinear behaviors.

The second application is derived from the work of Kursawe \cite{Kursawe1991}, and it has been defined in Eq. \ref{eq:kursawe}:
\begin{equation}
\label{eq:kursawe}
\begin{array}{lrl}
    k_1 (x_1,x_2,x_3) &=& \sum_{i=1}^2 \left[ -10 \exp \left( -0.2\sqrt{(x_i^2 + x_{i+1}^2)} \right) \right] \\[10pt]
    k_2 (x_1,x_2,x_3) &=& \sum_{i=1}^3 \left[ |x_i|^{0.8} + 5\sin (x_i^3) \right] \\[10pt]
    \mathrm{where} && -5 \leq x_i \leq 5.
\end{array}
\end{equation}
where $k_1(x_1,x_2,x_3)$ and $k_2(x_1,x_2,x_3)$ represent the EFPs of interest for the end-user, while $x_1$, $x_2$, and $x_3$ are the software-knobs of the application.
This application aims at validating the approach by increasing the complexity of the relation between the EFPs and the software-knobs.
In particular, $k_1(x_1,x_2,x_3)$ represents exponential functions, while $k_2(x_1,x_2,x_3)$  represents periodic functions.
These functions have been chosen to represent a large set of function types such as linear, nonlinear and exponential.

\subsubsection{Model Training and Validation}

\begin{figure*}
\centering
\begin{subfigure}{0.8\textwidth}
    \includegraphics[width=\textwidth]{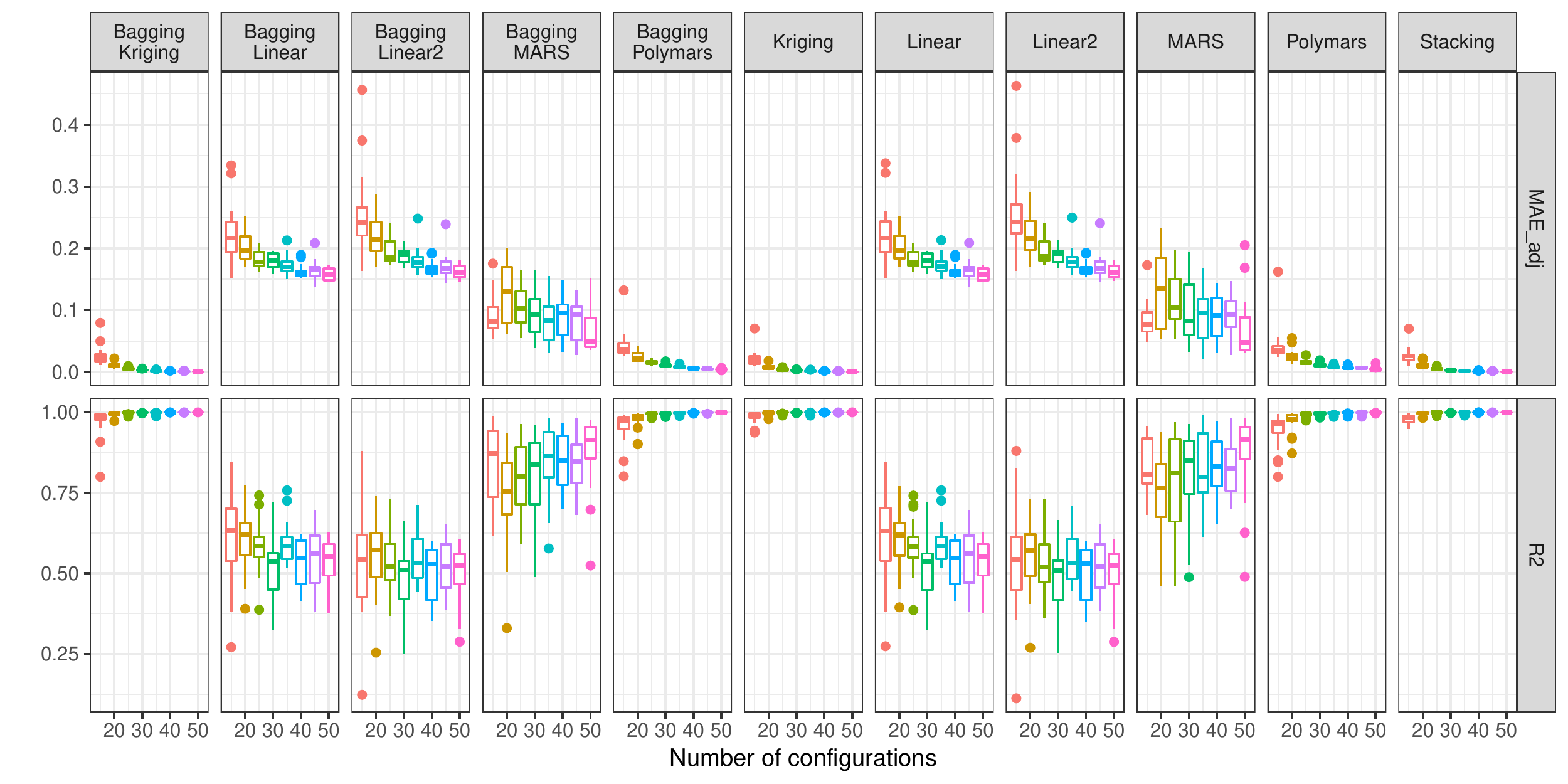}
    \caption{Validation of the $b_1(x,y)$ function.}
    \label{fig:results_binh_validation_1}
\end{subfigure}
\begin{subfigure}{0.8\textwidth}
    \includegraphics[width=\textwidth]{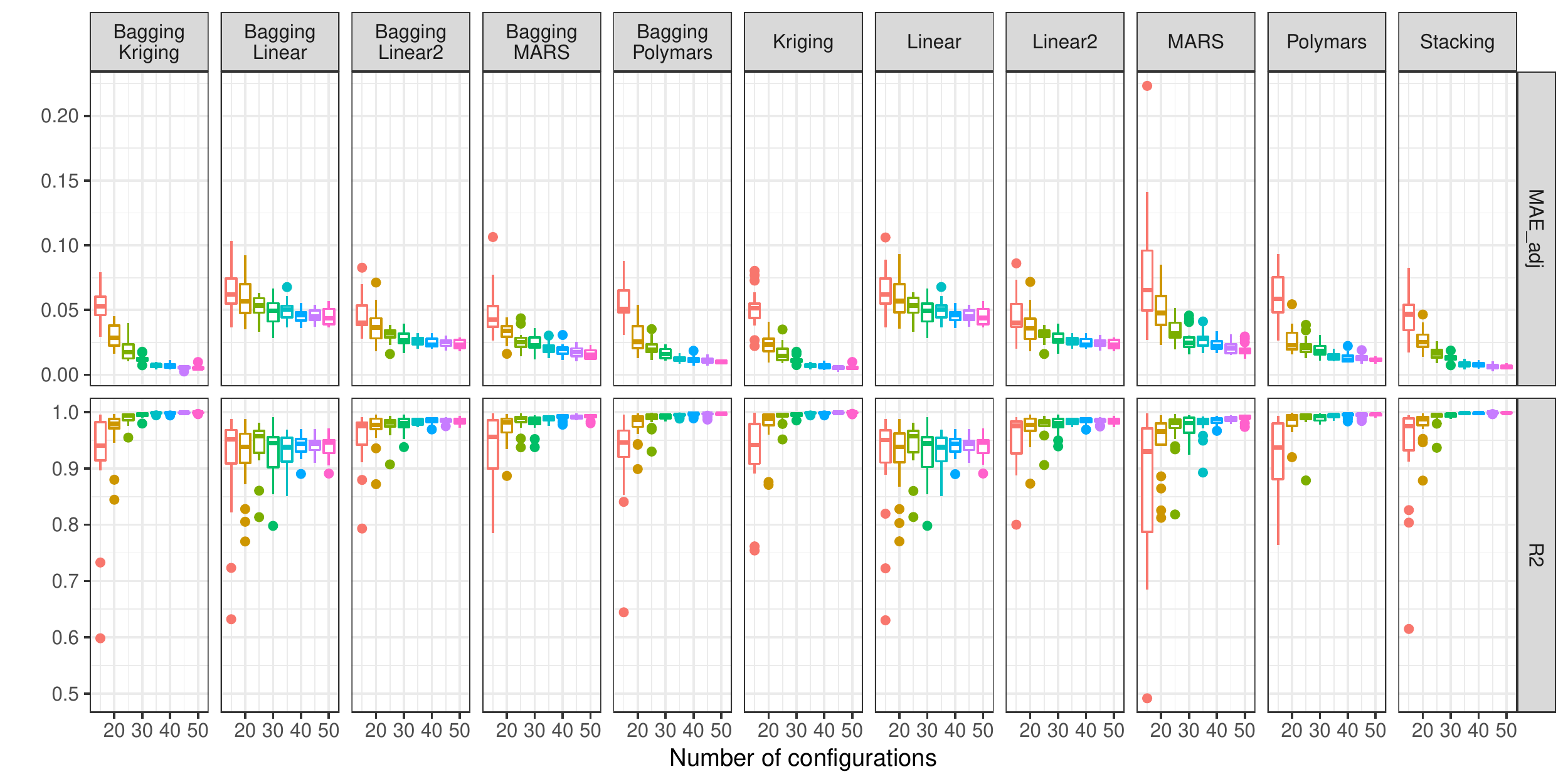}
    \caption{Validation of the $k_1(x_1,x_2,x_3)$ function.}
    \label{fig:results_kursawe_validation_1}
\end{subfigure}
\begin{subfigure}{0.8\textwidth}
    \includegraphics[width=\textwidth]{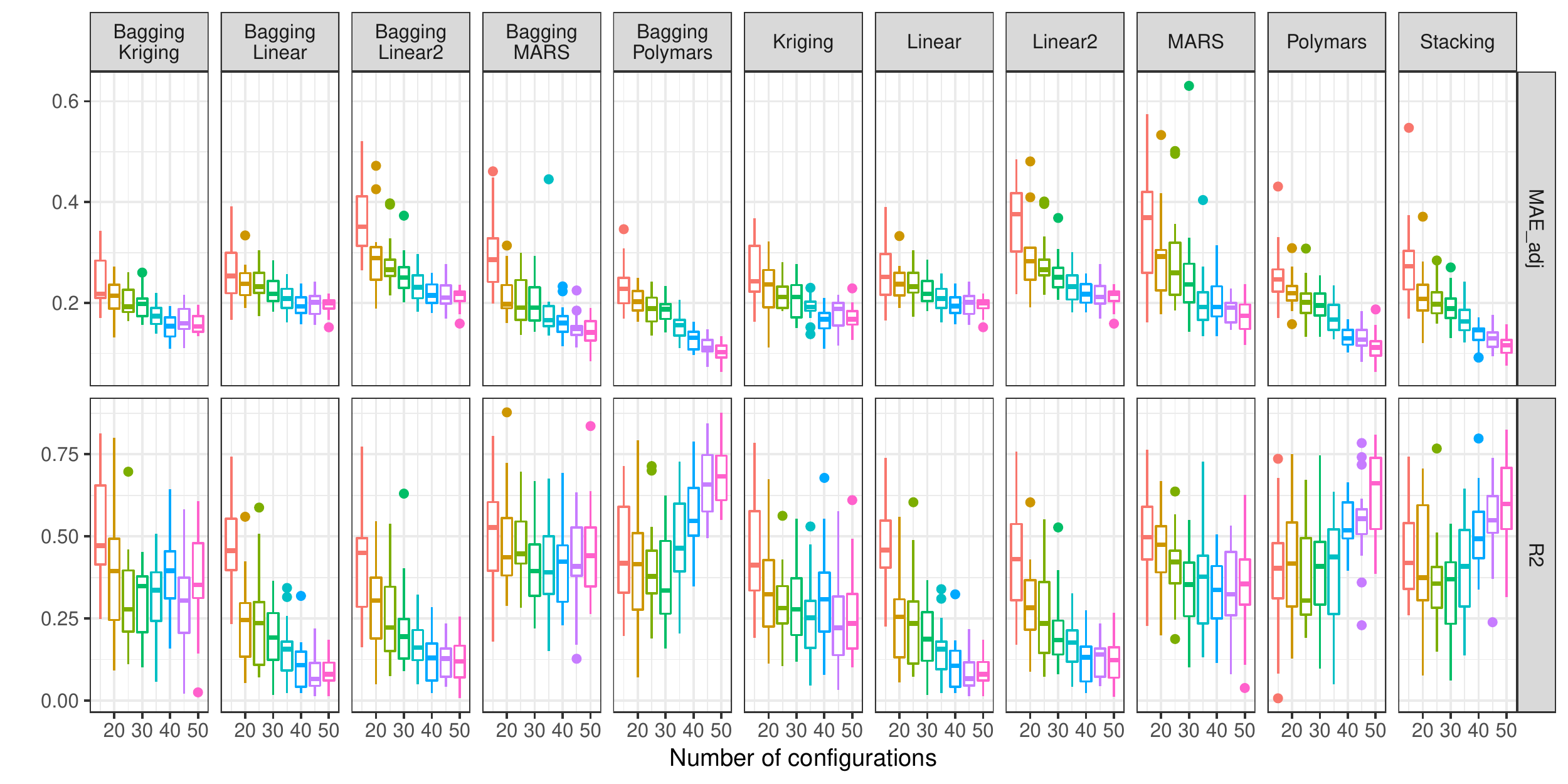}
    \caption{Validation of the $k_2(x_1,x_2,x_3)$ function.}
    \label{fig:results_kursawe_validation_2}
\end{subfigure}
\caption{Mean absolute error and correlation coefficient $R^2$ of the synthetic application EFPs models by varying the number of explored configurations during the DSE.}
\label{fig:validation_training}
\end{figure*}

In this section, we focus on evaluating the model training and validation by using a limited training set.
Figure \ref{fig:validation_training} shows the evaluation of the trained models for synthetic applications by varying the number of software-knob configurations evaluated in the DSE.
In particular, Figure \ref{fig:results_binh_validation_1} validates the trained models for the $b_1$ function, while Figure \ref{fig:results_kursawe_validation_1} and Figure \ref{fig:results_kursawe_validation_2} validate the trained models for the $k_1$ and the $k_2$ functions.
We omitted the validation results of the $b_2$ function because the underlying linear equation was learned well by almost all the models included in the proposed framework.
In Figure \ref{fig:validation_training}, the y-axis represents the model quality in terms of mean absolute error and correlation coefficient $R^2$, while the x-axis represents the number of software-knob configurations evaluated in the DSE.
For each x-value, we repeated the experiment 20 times and we have shown its distribution.

From the experimental results, we noticed a trend on the mean absolute error across all the model types and modelled functions: it decreases when we increase the number of explored software-knob configurations.
However, this is not true for the correlation coefficient.
Let us consider the $k_2$ EFPs of the Kursawe application.
We might notice how POLYMARS models to improve by increasing the number of training samples, while Kriging models seem to overfit on the training data, performing worse on the out-of-sample predictions.
Different model types behave differently in out-of-sample predictions according to the EFP underlying equation.
Let us focus on the MARS models across different EFPs.
We notice how according to the underlying equation of the target EFP, it behaves differently.
It is able to model the $k_1$ function, it struggles to model the $b_1$ function, while it fails to model the $k_2$ function.

If we consider all the EFPs, we might notice how at least one model type fares well in out-of-sample predictions in the validation set, except for the $k_2$ function.
If we focus on this EFP, it is possible to notice how the trained models struggle to explain the data, according to the correlation coefficient.
This is due to the periodic nature of the underlying equation and to the limited size of the training set.
Instead of learning the behaviour of the function, the trend is to settle along an average trend.
Therefore, the MAE decreases but the $R^2$ decreases as well.

\subsubsection{Model Selection}

\begin{figure}
\begin{subfigure}{0.5\textwidth}
    \includegraphics[width=\textwidth]{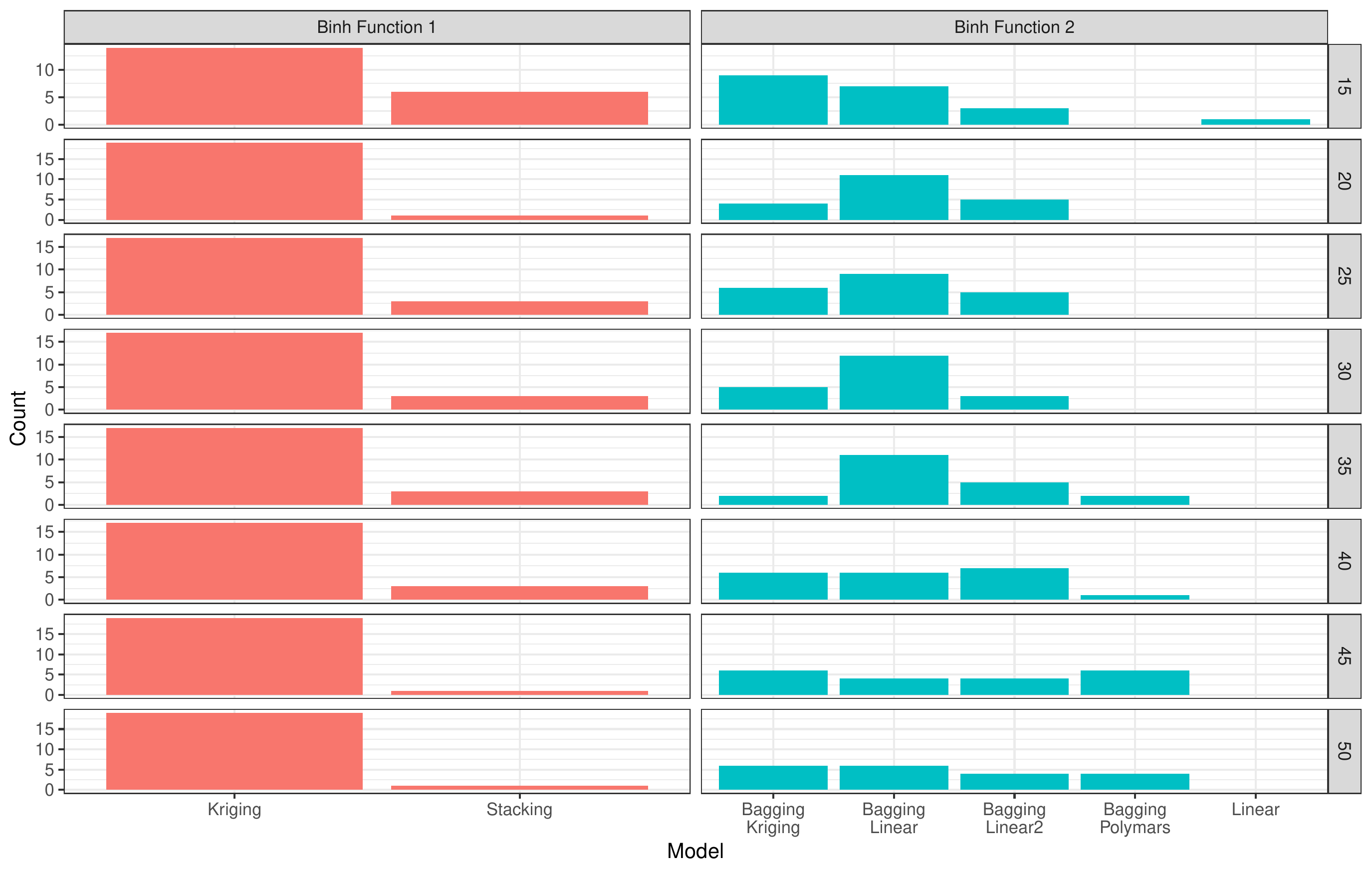}
    \caption{Binh synthetic application}
    \label{fig:selected_models_test4}
\end{subfigure}
\begin{subfigure}{0.5\textwidth}
    \includegraphics[width=\textwidth]{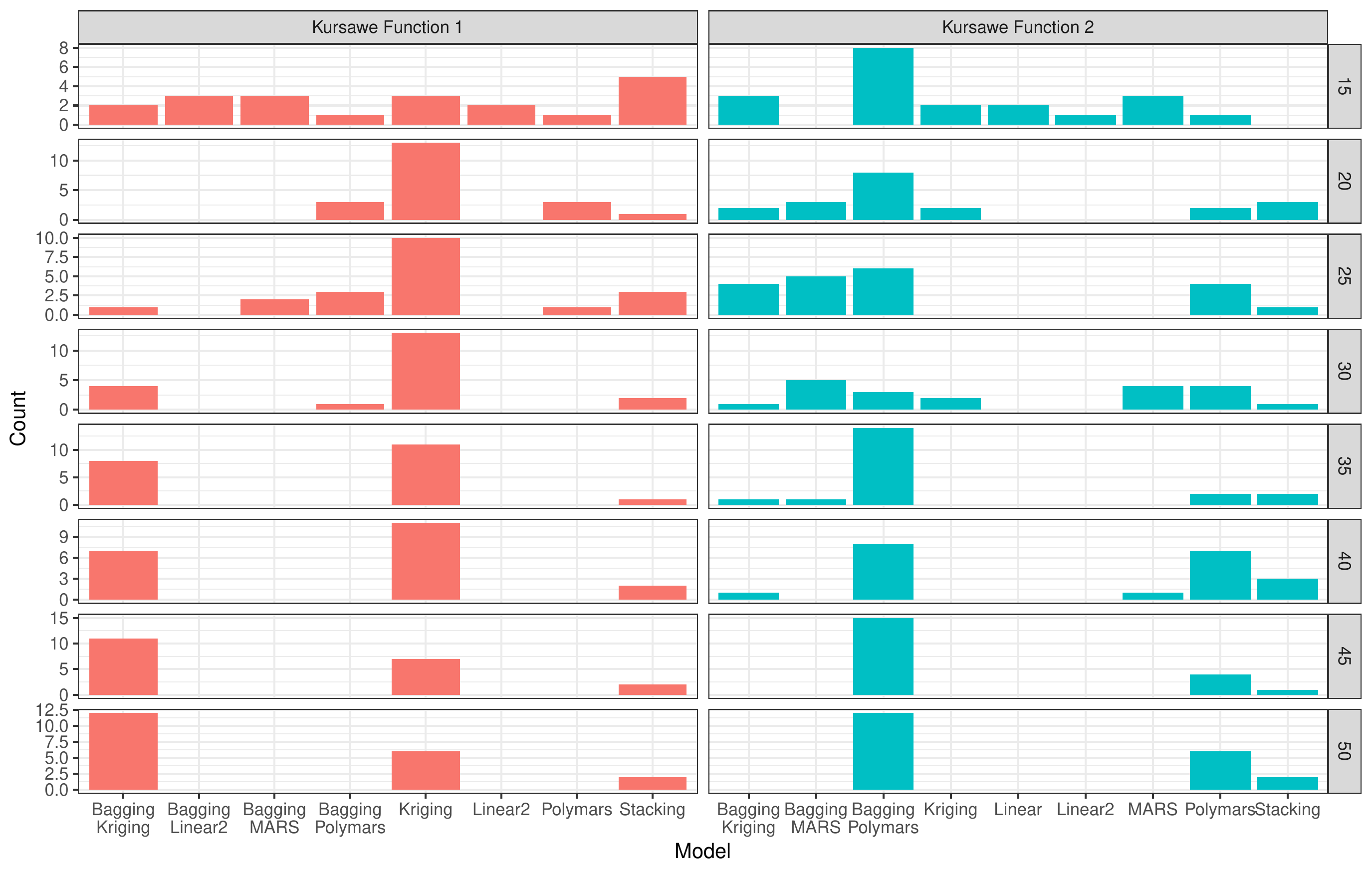}
    \caption{Kursawe synthetic application}
    \label{fig:selected_models_kursawe}
\end{subfigure}
\caption{Number of times that a model family is selected as the best one across the different EFPs of the two synthetic applications, by varying the number of software-knob configurations in the DSE.}
\label{fig:model_selection_validation}
\end{figure}

In this section, we evaluate the capability of the proposed framework to leverage different modelling techniques to learn the application knowledge at runtime.
Figure \ref{fig:model_selection_validation} shows the number of times that a model type is deemed as the best one, according to the methodology explained in Section \ref{ssec:model_validation_selection}.
For each EFP, we report the model selection according to the number of software-knob configurations explored during the DSE. Models not listed in Figure \ref{fig:model_selection_validation} have never been selected.

From the experimental results,  carried out on both synthetic applications, we notice how there is not a unique model always dominant with respect to the others. This confirms the importance of the proposed approach. 
The actual model selected strongly depends on the predicted function and on the values of the model selection parameters (i.e. $\epsilon_r$ and $\epsilon_m$).
If we focus on the models generated with a small training set, the learning module selects from a wider range of model families. On the opposite, with a larger number of samples (i.e. 50), the models more frequently selected are Kriging, Kriging bagged, POLYMARS bagged and stacking.
In all cases, the selected models provide good out-of-sample predictions because the cross-validation and testing error (as shown in Section \ref{ssec:framework_validation}) are comparable.

\subsubsection{Framework Validation}
\label{ssec:framework_validation}

\begin{figure}
\begin{subfigure}{0.45\textwidth}
    \includegraphics[width=\textwidth]{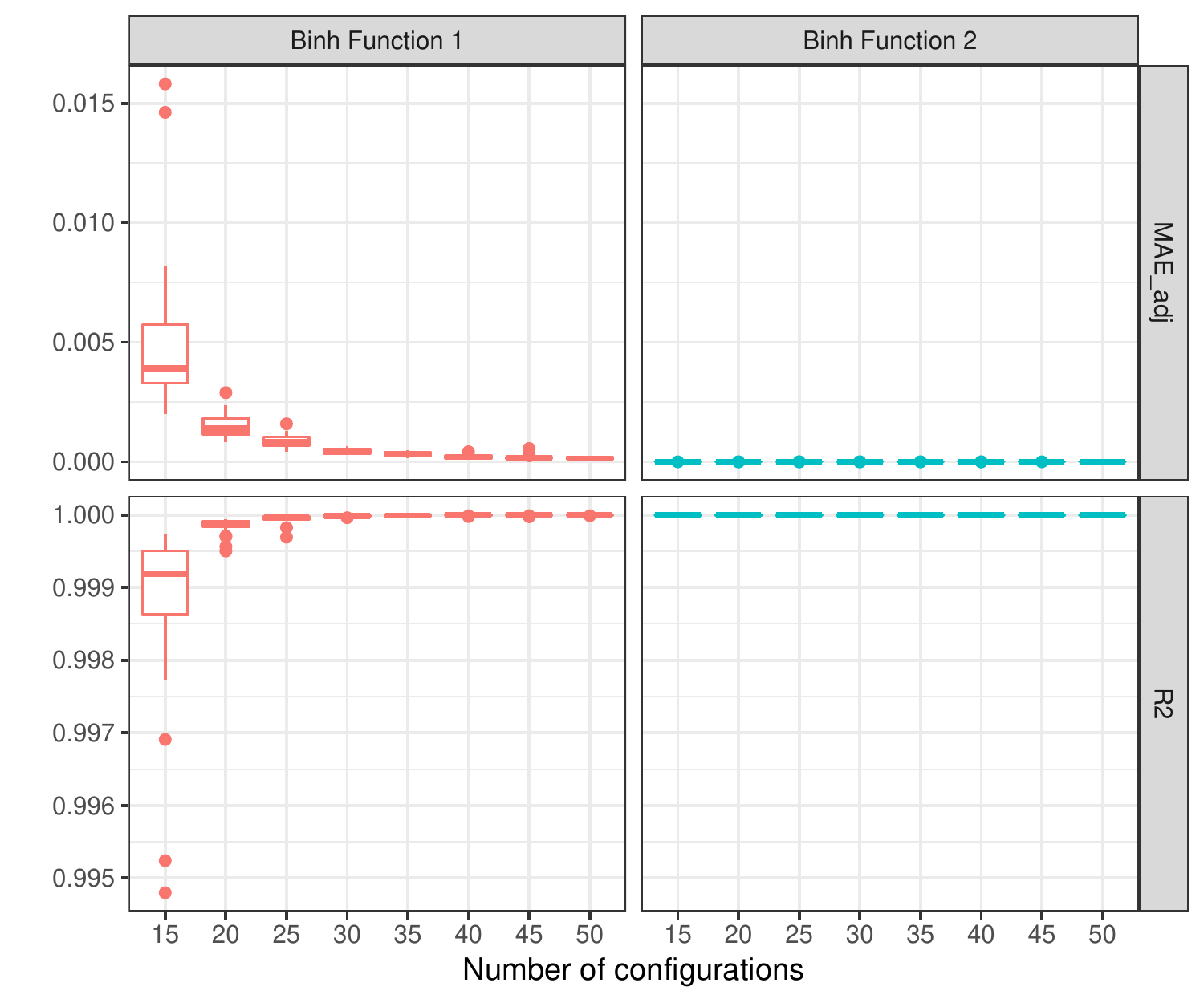}
    \caption{Binh synthetic application}
    \label{fig:binh_validation}
\end{subfigure}
\begin{subfigure}{0.45\textwidth}
    \includegraphics[width=\textwidth]{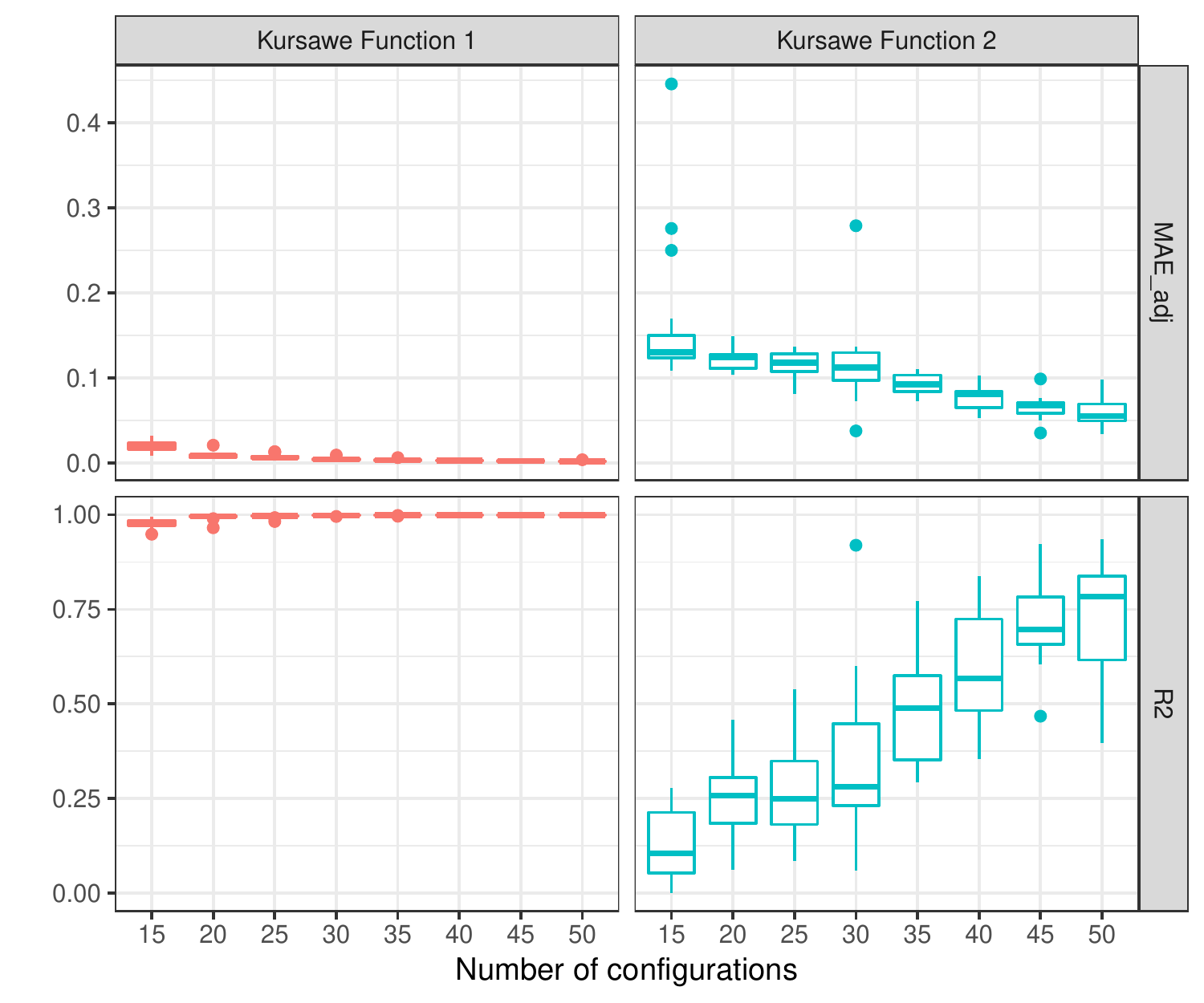}
    \caption{Kursawe synthetic application}
    \label{fig:kursawe_validation}
\end{subfigure}
\caption{Mean absolute error and correlation coefficient of the application knowledge generated by the proposed framework, by varying the number of software-knobs configurations.}
\label{fig:framework_validation}
\end{figure}

This section aims at evaluating the quality of the final output of the learning model: the application-knowledge.
In this experiment, we consider as input the application-knowledge given by the best model for the given run and we compare it with the underlying equation of the target EFP.
Figure \ref{fig:framework_validation} shows the experimental results in terms of prediction error ($MAE\_adj$) and $R^2$, across the whole software-knobs domain. Figure \ref{fig:binh_validation} refers to the Binh synthetic application, while Figure \ref{fig:kursawe_validation} refers to Kursawe synthetic application.
In both cases, the y-axis represents the quality of the application knowledge, while the x-axis represents the number of software-knob configurations used to compute the model.

From the experimental results, we can notice how the quality metrics are consistent with the validation of the model done in the training phase.
In particular, we can predict all the software-knob configurations of the design space within a $MAE\_adj$ of 4\% and a $R^2$ of 0.94, for all the EFPs of the synthetic applications, except for the $k_2$ EFP for the Kursawe application.
These results are coherent with the quality of the model determined in the validation phase, and used for the model selection.
This behavior is important because we can correctly judge the quality of the application-knowledge before broadcasting it to the application clients.
Therefore, the user is capable to manage the tradeoff between the learning phase duration and the quality of the result, by using the parameters $\epsilon_r$ and $\epsilon_m$.
Moreover, the proposed framework enable us to distribute the DSE to the application clients and to use an iterative refinement procedure to minimize the learning time of the model.

\subsection{The Molecular Docking Case Study}
\label{ssec:geodock} 
In this section, we validate the proposed approach on a real-life case study taken from the HPC world. 
First, we demonstrate the advantages obtained by adopting the proposed solution (Section \ref{ss:er:trace} and \ref{ss:er:prediciton}). Then, we evaluate some characteristics of the framework in terms of input feature clustering and scalability (Section \ref{ss:er:clustering} and \ref{ss:er:scalability}).

In a drug discovery process, molecular docking is one of the earliest tasks and it is performed \textit{in silico}.
Molecular docking is used to virtual screen a very large library of molecules, named \textit{ligands}, to find the ones with the strongest interaction with the binding site of a second molecule, named \textit{pocket}, to forward to later stages of the drug discovery process \cite{beccari2013ligen}.
The complexity of this task is not only due to the huge number of ligands to evaluate, but also to the number of degrees of freedom in the evaluation of the ligand-pocket interaction.
In particular, it is possible to alter the shape of the molecule, without altering its chemical properties, by rotating a subset of bonds between the atoms of a ligand, named \textit{rotamers}.

In this experiment, we focus on a geometric docking kernel, part of the LiGen Dock application \cite{beato2013use}.
Due to the complexity of evaluating the chemical interaction of a pocket-ligand pair, this kernel considers only geometrical information and it is used to filter out the ligands unable to fit in the target pocket.
The application exposes two software-knobs that generate quality-throughput tradeoffs by reducing the number of alternative poses evaluated for each rotamer of the ligand.
The end-users typically belong to pharmaceutical companies that rent the resources of an HPC infrastructure, to evaluate a chemical library by running a typical batch job.
Therefore, the end-users are interested in time-to-solution and on the quality of the elaboration, defined as the number of evaluated poses.

The goal of this experiment is to assess the benefits of the proposed framework where the throughput is heavily input-dependent.
The time spent on evaluating a pocket-ligand pair depends on the number of atoms and rotamers of the ligand, and by the geometrical properties of the pocket that are difficult to represent numerically.
This heavy input dependency is perceived as a significant noise when measuring the execution time for a given software-knob configuration across several ligands.
Given that the pocket remains the same for the entire screening process, the proposed approach aims at learning the effect of the target pocket on the relation between software-knobs and the execution time at the production phase.
Every time the pocket changes, we need to learn again the application-knowledge, however, 1) a pocket is seldom re-evaluated, 2) the time spent on evaluating the library is several orders of magnitude larger than the learning time, 3) the application-knowledge is tailored to the actual input, and 4) we exploit the parallelism of the production system (composed of several HPC nodes).

Given that the later stages of the drug discovery process require an expensive cost to execute the \textit{in-vitro} and \textit{in-vivo} tests, the reproducibility of the experiment is a domain-requirement.
Therefore, once we have obtained the application-knowledge, we restart to evaluate the chemical library with the configuration that maximizes the quality, while respecting the time-to-solution constraint adjusted by the time spent during the learning phase.
In the following experiments, we used a library of $113k$ ligands, where each ligand has a number of atoms between $28$ and $153$ and a number of rotamers between $2$ and $53$.
Moreover, we use six pockets (\textit{1b9v}, \textit{1c1b}, \textit{1cvu}, \textit{1cx2}, \textit{1dh3} and \textit{1fm9}) from the RCSB Protein Databank (PDB) \cite{PDB}.

In the next sections, we validate the proposed approach by using different experiments. First, we show the typical execution traces where the online learning approach has been employed. Second, we validate the approach by simulating a virtual screening campaign. 
Third, we demonstrate the advantages introduced by the input feature clustering module to reduce the prediction variability. Finally, we use this case study to evaluate the scalability in terms of the time-to-learn for the proposed approach.

\begin{figure*}
    \centering
        \begin{subfigure}{0.49\columnwidth}
        \centering
        \includegraphics[width=1\linewidth]{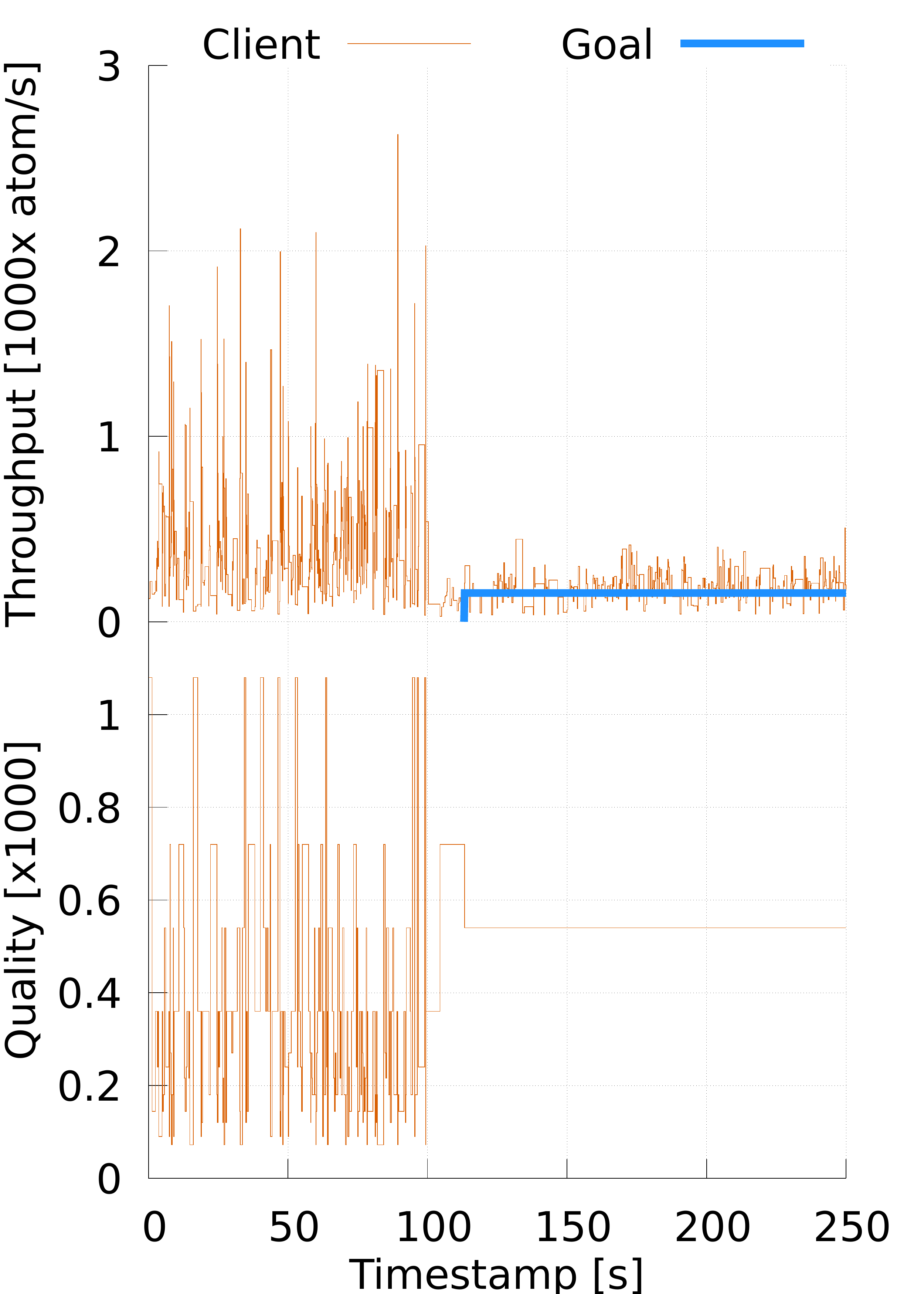}
        \caption{MPI process 1}
        \label{fig:pocket_1}
    \end{subfigure}
    \begin{subfigure}{0.49\columnwidth}
        \centering
        \includegraphics[width=1\linewidth]{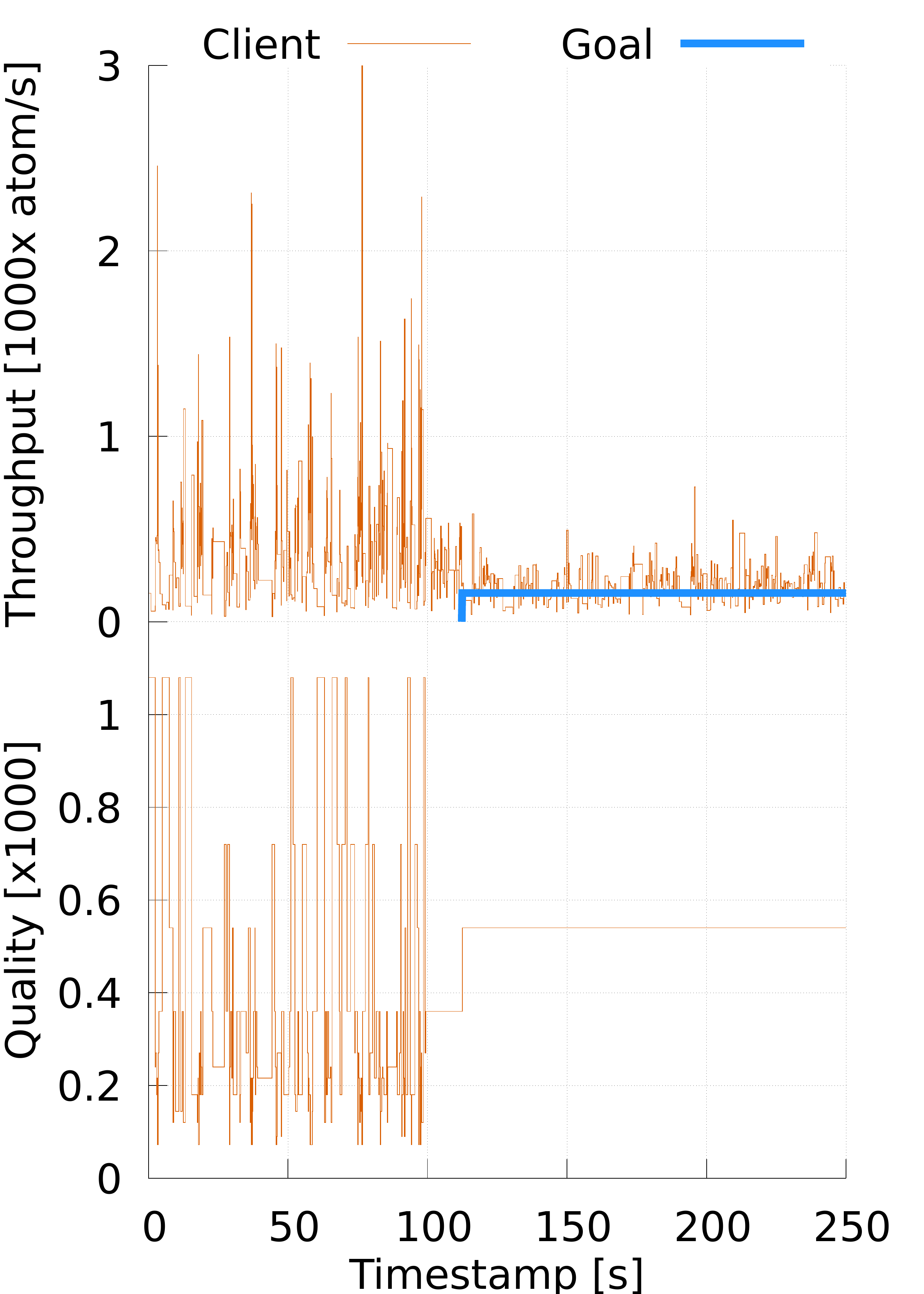}
        \caption{MPI process 2}
        \label{fig:pocket_2}
    \end{subfigure}
    \begin{subfigure}{0.49\columnwidth}
        \centering
        \includegraphics[width=1\linewidth]{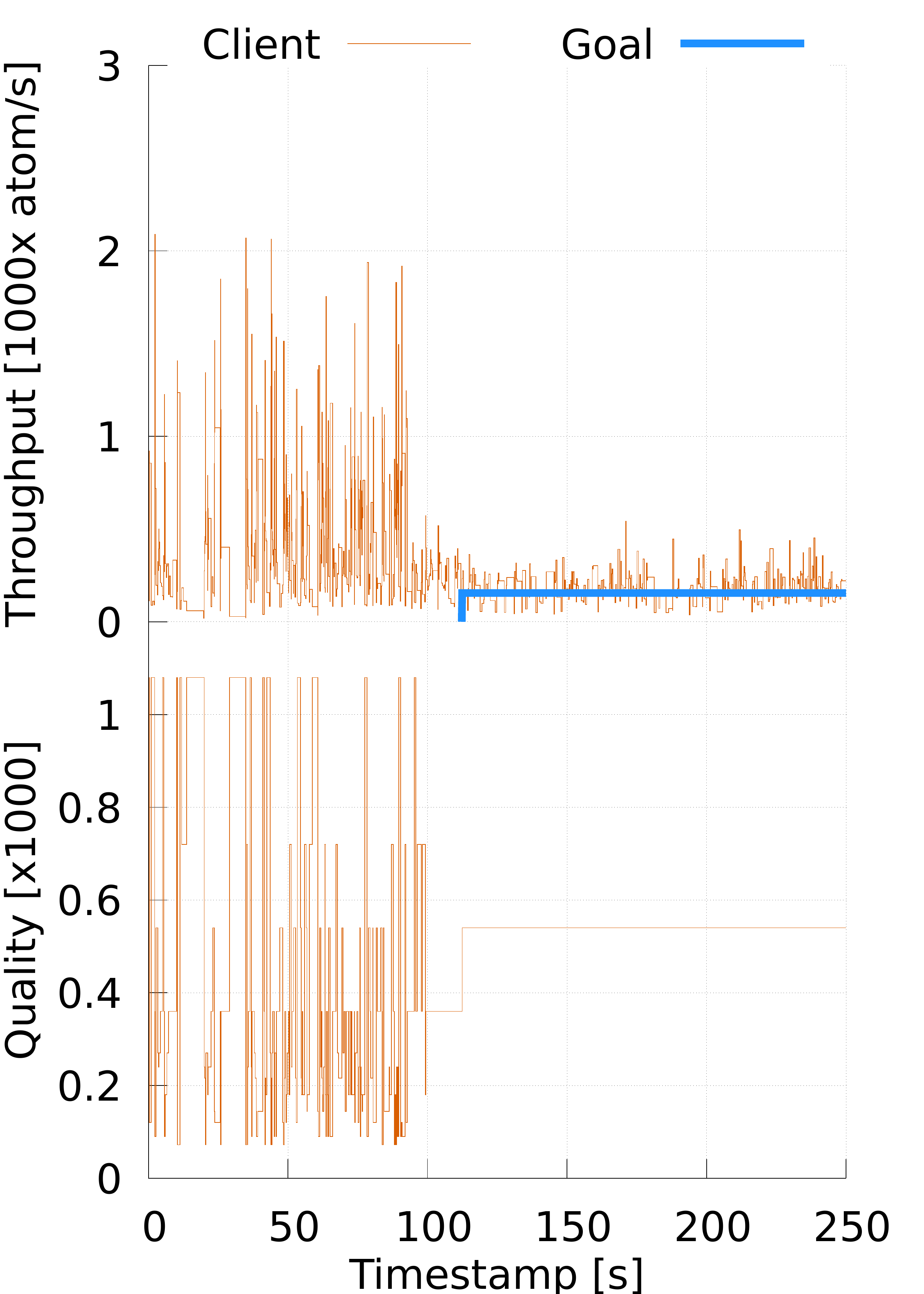}
        \caption{MPI process 3}
        \label{fig:pocket_3}
    \end{subfigure}
    \begin{subfigure}{0.49\columnwidth}
        \centering
        \includegraphics[width=1\linewidth]{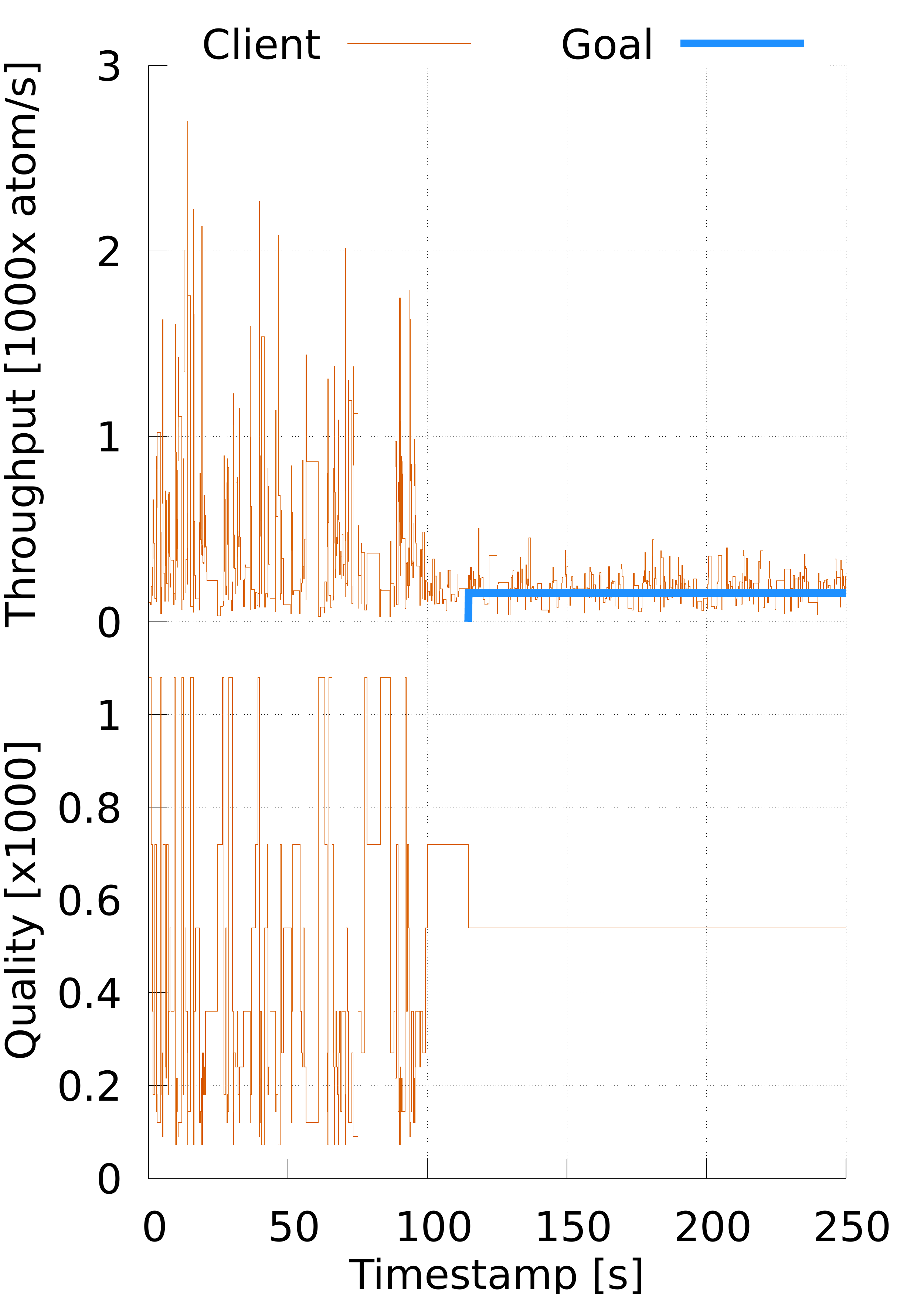}
        \caption{MPI process 4}
        \label{fig:pocket_4}
    \end{subfigure}
    \caption{Execution trace of the docking application learning phase by using 16 MPI process and targeting the pocket $1b9v$. For $4$ out of $16$ MPI processes, we show the throughput for evaluating the pocket-ligand pair and the quality over the firsts $250$ seconds of a longer virtual screening process.
    \label{fig:geodock_execution_trace}}
\end{figure*}

\subsubsection{Execution Trace}
\label{ss:er:trace}
Figure \ref{fig:geodock_execution_trace} shows an execution trace of the docking application running on $16$ MPI processes on the pocket \textit{1b9v}.
Figure \ref{fig:geodock_execution_trace} represents the application behavior in terms of the throughput for evaluating the pocket-ligand pair and the quality of the results over the firsts $250$ seconds of a large experiment.
Each sub-figure shows the EFP behavior of $4$ out of $16$ MPI processes, as representative behaviour. 
The length of the learning phases is determined by the model convergence time and on the input characteristics.
It is possible to notice that the length is almost the same for all the client's thanks to the configuration distribution performed by the remote application handler. 
After the learning phase, the goal value has been set by computing the average throughput required to process the entire target ligand database given the target time-to-solution (experiment target budget).
After the initial exploration of the design space, the application settles with a software-knob configuration that is the same for all the clients.
This is because all clients are part of the same virtual screening experiment.
By varying the current inputs (i.e. the target pocket and the target ligand database) or the time-to-solution constraint, the autotuning process would lead to a different configuration.

\subsubsection{Prediction Accuracy}
\label{ss:er:prediciton}
\begin{figure}
    \centering
    \includegraphics[width=\linewidth]{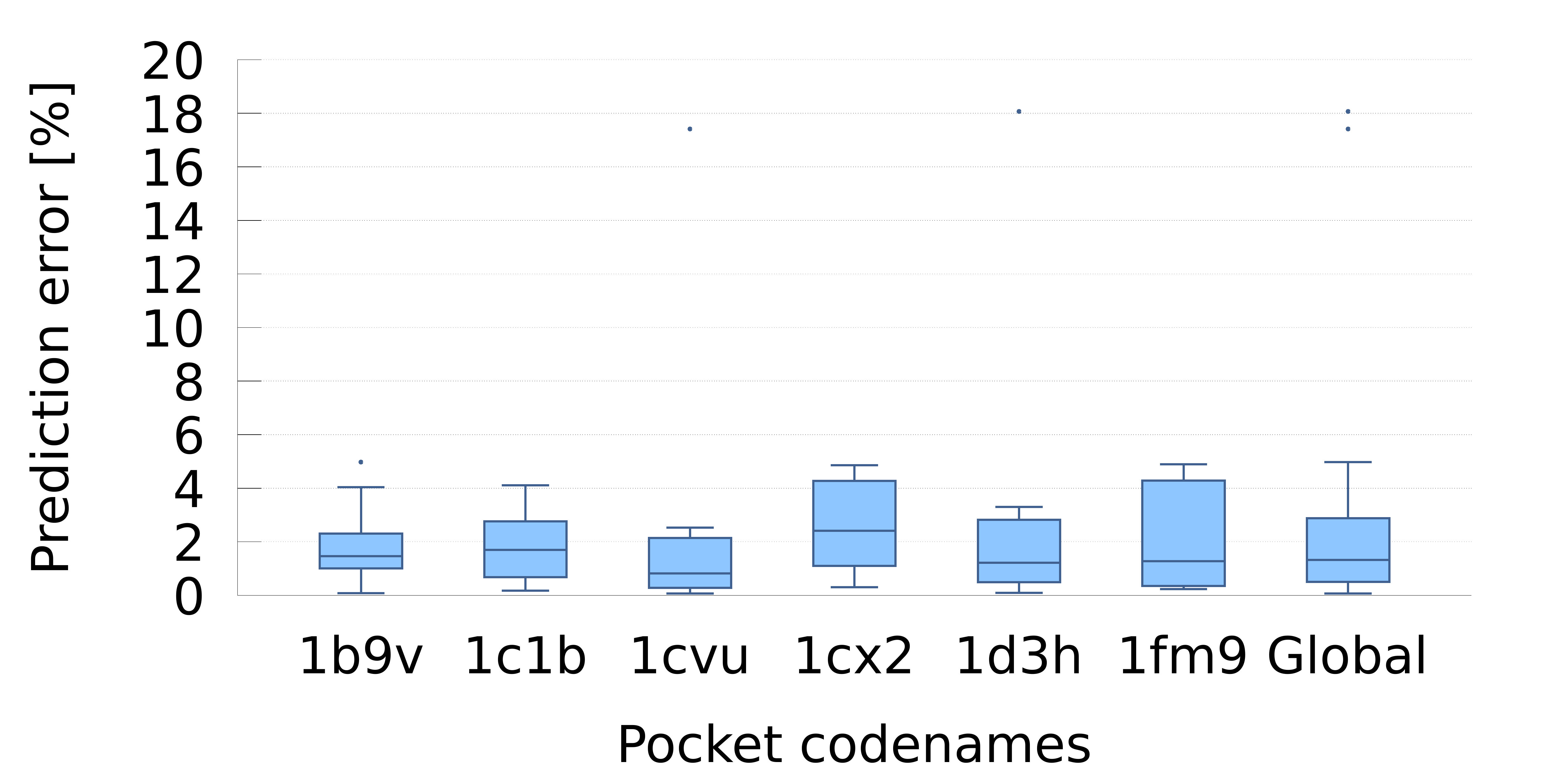}
    \caption{Distribution of the prediction error in percentage, grouped by different target pockets.}
    \label{fig:geodock_error}
\end{figure}

To validate the proposed approach, we run an experimental campaign with a library of $8k$ ligands, randomly sampled from $113k$ ligands, targeting six different pockets.
In particular, for each pocket we repeated the experiment ten times, reporting the prediction error distribution
(see Figure \ref{fig:geodock_error}).
For the learning phase, we observed each software-knob configuration in the DoE with $400$ different ligands.
Figure \ref{fig:geodock_error} shows that a large fraction of the time-to-solution errors are within $5\%$.
In this case, the proposed approach accurately estimates the time-to-solution for the current inputs, maximizing the quality of the results given the time budget.
In this experiment, we found that Kriging, MARS bagged, and stacked models are the top three models according to $MAE\_adj$, the actual selected model varied across the experiments.

\subsubsection{Input Feature Clustering}
\label{ss:er:clustering}

\begin{figure}
    \centering
    \includegraphics[width=\linewidth]{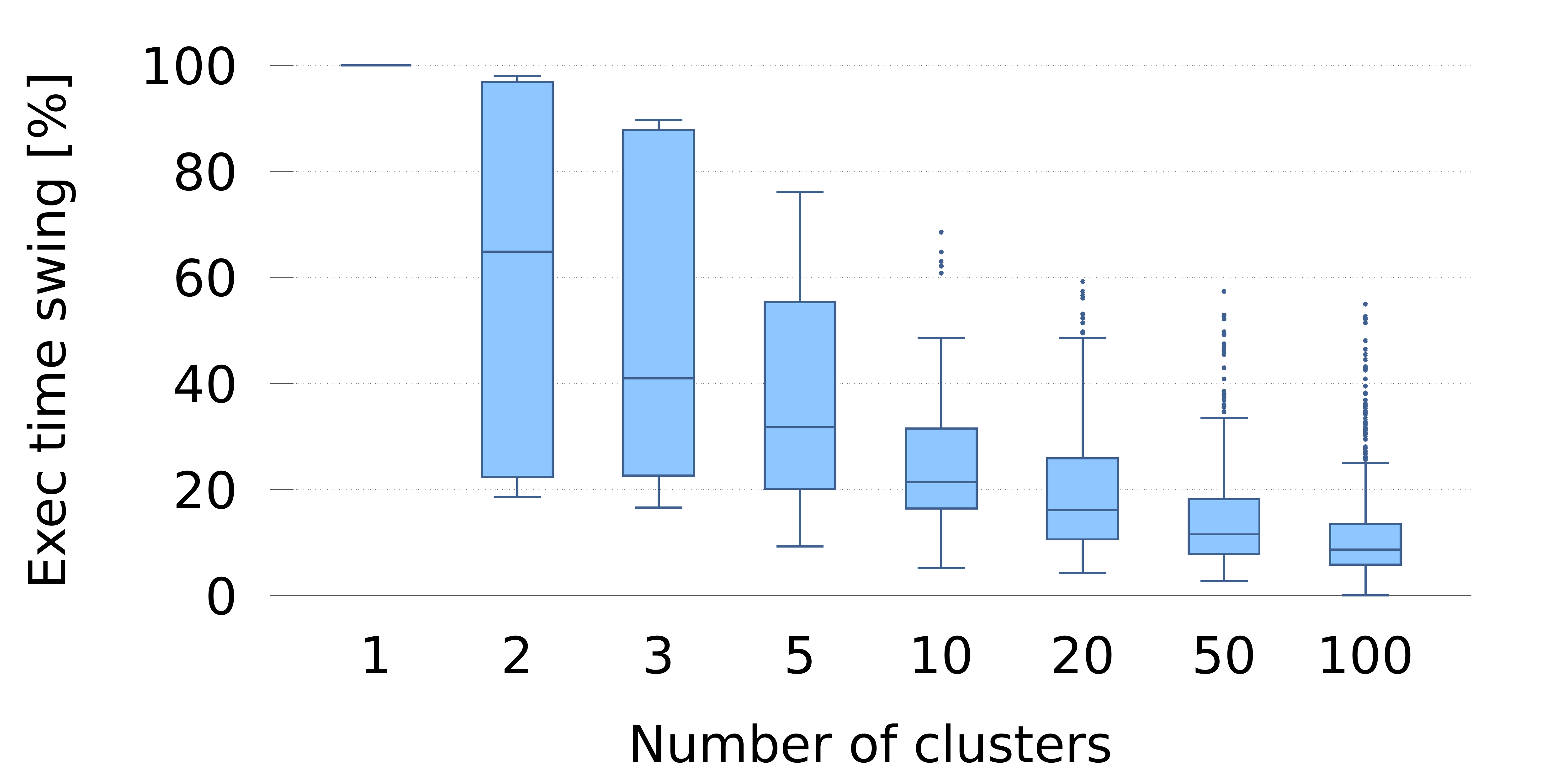}
    \caption{Distribution of the normalized execution time swing in each cluster by varying the number of clusters}
    \label{fig:geodock_clustering}
\end{figure}

In this subsection, we demonstrate the benefits introduced by the input feature clustering module to reduce the prediction error variability and the training set dependency. 
In particular, Figure \ref{fig:geodock_clustering} shows the effects obtained by increasing the number of input feature clusters on the prediction accuracy. 
The clusters have been determined by using the K-means algorithm over the number of atoms of the molecule and the number of rotamers.
The prediction accuracy has been measured by considering the variability of the docking time of each molecule represented by the same input feature cluster. This value has been normalized to the one obtained without considering the input features (i.e. one single cluster). 
Increasing this value, the execution time swing reduces on average and the distribution of the data becomes tighter. Despite a rapid initial reduction, the swing does not lead to zero because there are some input characteristics not completely captured by the data features used for the clustering (e.g. the geometry of the pocket and the ligand).
Increasing the number of clusters, on one side, we reduce the execution time swing, while, on the other side, we are increasing the amount of memory needed to store the mARGOt OP list.

The reduction of the EFP variability in each cluster is even more important when the data used during the training phase are not distributed as for the whole data-set for the experiment. Indeed, considering a single cluster (no input features) means that the average behaviour learned during the training phase will be the same also for other data. This happens either if the EFPs are not dependent on the input data, or if the input data are on average the same as those used for the training.
If we are not in the previous cases, having more clusters enables a better prediction process of the EFP values.

\subsubsection{Scalability Analysis}
\label{ss:er:scalability}

\begin{figure}
    \centering
    \includegraphics[width=\linewidth]{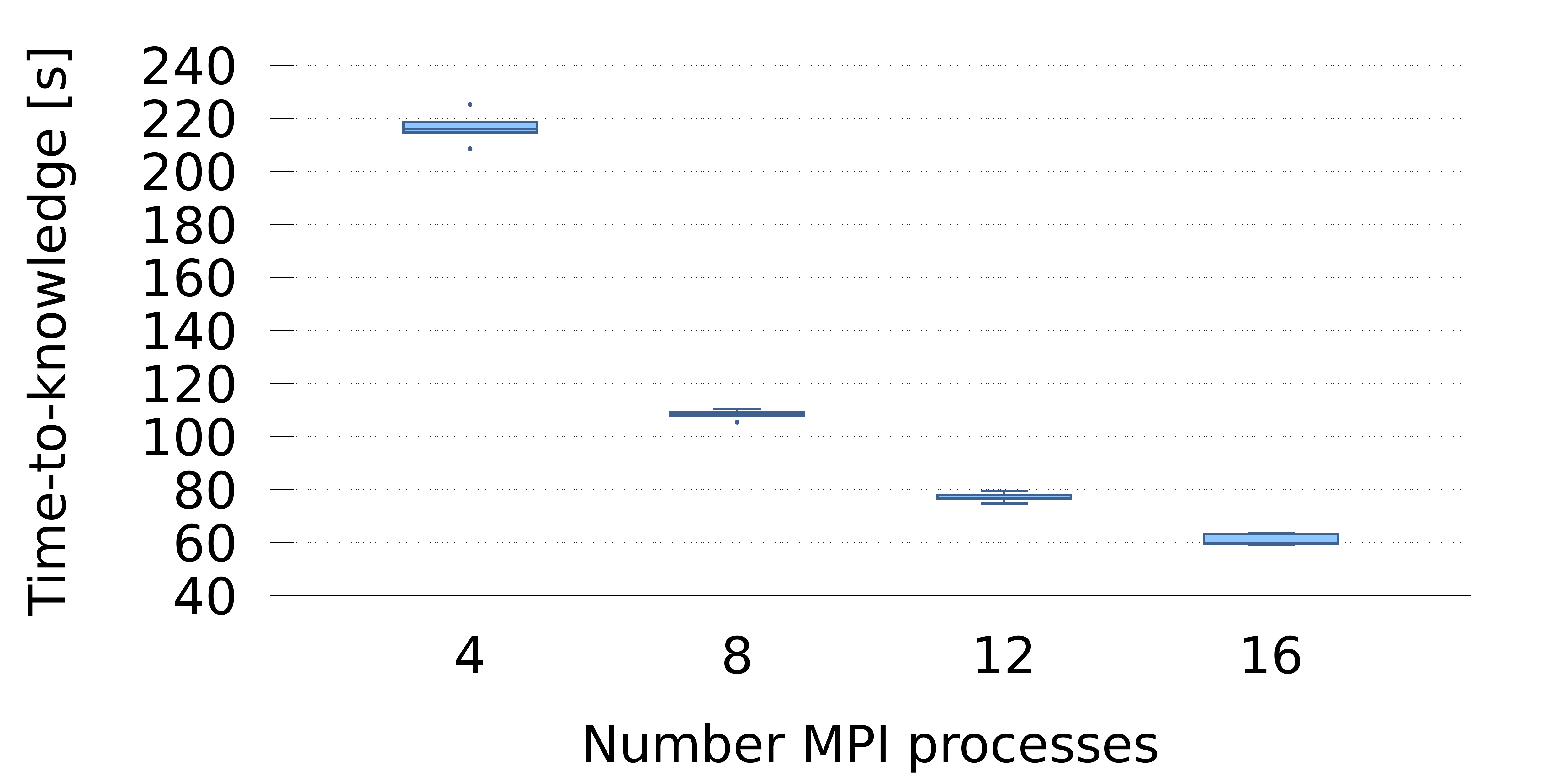}
    \caption{Distribution of the time spent for learning the application-knowledge by varying the number of MPI processes}
    \label{fig:geodock_learning}
\end{figure}

This experiment addresses the scalability of the proposed approach in terms of time-to-learn.
In particular, we show how the time requested to generate the application-knowledge decreases according to the number of clients that contribute to the exploration phase.
Figure \ref{fig:geodock_learning} shows the distribution of the time required by an application-client to receive the application-knowledge.
For each number of MPI processes, we repeated the experiment five times.
From the results, the overhead is almost inversely proportional to the number of clients. Indeed, the time-to-knowledge almost halved when doubling the number of MPI processes. 
In particular, the proposed framework required about $10$ seconds for the training and validation of all models.
This time is short enough when considering the number of models we are training and the target use case.
The time needed by the training, validation and selection phase is constant regardless of the number of clients.

\section{Conclusions}
\label{sec:conclusion}

This paper proposed an online autotuning framework to learn the relation between software-knobs, extra-functional properties and input features at the production phase, in a distributed way.
To minimize the learning time, the framework is based on two strategies.
On one side, it uses ensemble models to boost the prediction capabilities of the base models. On the other side, it uses an iterative approach for sampling the design space until the computed models reach the target quality.

Experimental results on synthetic applications and on a real-world case study, demonstrate how there is\textit{no-free-lunch}: it does not exist one model to fit all the cases.
This result confirms the main goal of the proposed approach: the goal was not to  compare different modelling techniques, but to provide a framework exploiting them to learn the application-knowledge at the runtime.

\bibliographystyle{abbrv}
\bibliography{bibliography}

\begin{thebibliography}{10}

\bibitem{ansel2014opentuner}
J.~Ansel~et al.
\newblock Opentuner: An extensible framework for program autotuning.
\newblock In {\em PACT}. IEEE, 2014.

\bibitem{baek2010green}
W.~Baek and T.~M. Chilimbi.
\newblock Green: a framework for supporting energy-conscious programming using
  controlled approximation.
\newblock In {\em ACM Sigplan Notices}, volume~45, pages 198--209. ACM, 2010.

\bibitem{AutotuningInHPC}
P.~Balaprakash, J.~Dongarra, T.~Gamblin, M.~Hall, J.~K. Hollingsworth,
  B.~Norris, and R.~Vuduc.
\newblock Autotuning in high-performance computing applications.
\newblock {\em Proceedings of the IEEE}, 106(11):2068--2083, Nov 2018.

\bibitem{beato2013use}
C.~Beato~et al.
\newblock Use of experimental design to optimize docking performance: The case
  of ligendock, the docking module of ligen, a new de novo design program,
  2013.

\bibitem{beccari2013ligen}
A.~R. Beccari, C.~Cavazzoni, C.~Beato, and G.~Costantino.
\newblock Ligen: a high performance workflow for chemistry driven de novo
  design, 2013.

\bibitem{PDB}
H.~M. Berman, J.~Westbrook, Z.~Feng, G.~Gilliland, T.~N. Bhat, H.~Weissig,
  I.~N. Shindyalov, and P.~E. Bourne.
\newblock The protein data bank.
\newblock {\em Nucleic Acids Res}, 28:235--242, 2000.

\bibitem{Binh1999}
T.~T. Binh.
\newblock {A Multiobjective Evolutionary Algorithm - The Study Cases}.
\newblock {\em INSTITUTE FOR AUTOMATION AND COMMUNICATION}, 1999.

\bibitem{Breiman1996_bagging}
L.~Breiman.
\newblock {Bagging predictors}.
\newblock {\em Machine Learning}, 1996.

\bibitem{Breiman1996}
L.~Breiman.
\newblock {Stacked regressions}.
\newblock {\em Machine Learning}, 24, 1996.

\bibitem{christen2011patus}
M.~Christen, O.~Schenk, and H.~Burkhart.
\newblock Patus: A code generation and autotuning framework for parallel
  iterative stencil computations on modern microarchitectures.
\newblock In {\em Parallel \& Distributed Processing Symposium (IPDPS), 2011
  IEEE International}, pages 676--687. IEEE, 2011.

\bibitem{ding2015autotuning}
Y.~Ding~et al.
\newblock Autotuning algorithmic choice for input sensitivity.
\newblock In {\em ACM SIGPLAN Notices}, volume~50. ACM, 2015.

\bibitem{dorn2017automatically}
J.~Dorn, J.~Lacomis, W.~Weimer, and S.~Forrest.
\newblock Automatically exploring tradeoffs between software output fidelity
  and energy costs.
\newblock {\em IEEE Transactions on Software Engineering}, 2017.

\bibitem{Dupuy2015}
D.~Dupuy~et al.
\newblock {DiceDesign and DiceEval: Two R Packages for Design and Analysis of
  Computer Experiments}.
\newblock {\em Journal of Statistical Software}, 2015.

\bibitem{esmaeilzadeh2011dark}
H.~Esmaeilzadeh~et al.
\newblock Dark silicon and the end of multicore scaling.
\newblock In {\em ISCA}. IEEE, 2011.

\bibitem{ester1996density}
M.~Ester, H.-P. Kriegel, J.~Sander, X.~Xu, et~al.
\newblock A density-based algorithm for discovering clusters in large spatial
  databases with noise.
\newblock In {\em Kdd}, volume~96, pages 226--231, 1996.

\bibitem{Everitt2010}
B.~Everitt~et al.
\newblock {\em {The Cambridge dictionary of statistics}}.
\newblock 2010.

\bibitem{Friedman1991}
J.~H. Friedman.
\newblock {Multivariate Adaptive Regression Splines}.
\newblock {\em The Annals of Statistics}, 1991.

\bibitem{frigo2005design}
M.~Frigo and S.~G. Johnson.
\newblock The design and implementation of fftw3.
\newblock {\em Proceedings of the IEEE}, 93(2):216--231, 2005.

\bibitem{gadioli2018margot}
D.~Gadioli, E.~Vitali, G.~Palermo, and C.~Silvano.
\newblock margot: a dynamic autotuning framework for self-aware approximate
  computing.
\newblock {\em IEEE Transactions on Computers}, 2018.

\bibitem{guo2003bayesian}
H.~Guo.
\newblock A bayesian approach for automatic algorithm selection.
\newblock In {\em Proceedings of the International Joint Conference on
  Artificial Intelligence (IJCAI03), Workshop on AI and Autonomic Computing,
  Acapulco, Mexico}, pages 1--5, 2003.

\bibitem{hartigan1979algorithm}
J.~A. Hartigan and M.~A. Wong.
\newblock Algorithm as 136: A k-means clustering algorithm.
\newblock {\em Journal of the Royal Statistical Society. Series C (Applied
  Statistics)}, 28(1):100--108, 1979.

\bibitem{hoffmann2009using}
H.~Hoffmann~et al.
\newblock Using code perforation to improve performance, reduce energy
  consumption, and respond to failures.
\newblock 2009.

\bibitem{hoffmann2011dynamic}
H.~Hoffmann~et al.
\newblock Dynamic knobs for responsive power-aware computing.
\newblock In {\em ACM SIGPLAN Notices}. ACM, 2011.

\bibitem{kamil2012productive}
S.~A. Kamil.
\newblock {\em Productive high performance parallel programming with auto-tuned
  domain-specific embedded languages}.
\newblock University of California, Berkeley, 2012.

\bibitem{kephart2003vision}
J.~Kephart~et al.
\newblock The vision of autonomic computing.
\newblock {\em Computer 2003}, 36, 2003.

\bibitem{POLYMARS}
C.~Kooperberg.
\newblock {\em polspline: Polynomial Spline Routines}, 2018.
\newblock R package version 1.1.13.

\bibitem{Kursawe1991}
F.~Kursawe.
\newblock {A variant of evolution strategies for vector optimization}.
\newblock In {\em Lecture Notes in Computer Science}, 1991.

\bibitem{laurenzano2016input}
M.~A. Laurenzano~et al.
\newblock Input responsiveness: using canary inputs to dynamically steer
  approximation.
\newblock {\em ACM SIGPLAN Notices}, 2016.

\bibitem{mahdavi2017systematic}
S.~Mahdavi-Hezavehi, V.~H. Durelli, D.~Weyns, and P.~Avgeriou.
\newblock A systematic literature review on methods that handle multiple
  quality attributes in architecture-based self-adaptive systems.
\newblock {\em Information and Software Technology}, 90:1--26, 2017.

\bibitem{miguel2016anytime}
J.~S. Miguel~et al.
\newblock The anytime automaton.
\newblock In {\em ACM SIGARCH Computer Architecture News}. IEEE Press, 2016.

\bibitem{misailovic2013parallelizing}
S.~Misailovic, D.~Kim, and M.~Rinard.
\newblock Parallelizing sequential programs with statistical accuracy tests.
\newblock {\em ACM Transactions on Embedded Computing Systems (TECS)},
  12(2s):88, 2013.

\bibitem{mittal2016survey}
S.~Mittal.
\newblock A survey of techniques for approximate computing.
\newblock {\em ACM Computing Surveys (CSUR)}, 48(4):62, 2016.

\bibitem{montgomery2017design}
D.~C. Montgomery.
\newblock {\em Design and analysis of experiments}.
\newblock John wiley \& sons, 2017.

\bibitem{nugteren2015cltune}
C.~Nugteren and V.~Codreanu.
\newblock Cltune: A generic auto-tuner for opencl kernels.
\newblock In {\em Embedded Multicore/Many-core Systems-on-Chip (MCSoC), 2015
  IEEE 9th International Symposium on}, pages 195--202. IEEE, 2015.

\bibitem{quadprog}
S.~original by Berwin A. Turlach R port~by Andreas
  Weingessel~<Andreas.Weingessel@ci.tuwien.ac.at>.
\newblock {\em quadprog: Functions to solve Quadratic Programming Problems.},
  2013.
\newblock R package version 1.5-5.

\bibitem{MARS}
S.~original by Trevor Hastie \& Robert Tibshirani. Original R port~by
  Friedrich~Leisch, K.~Hornik, and B.~D. Ripley.
\newblock {\em mda: Mixture and Flexible Discriminant Analysis}, 2017.
\newblock R package version 0.4-10.

\bibitem{puschel2004spiral}
M.~P{\"u}schel, J.~M. Moura, B.~Singer, J.~Xiong, J.~Johnson, D.~Padua,
  M.~Veloso, and R.~W. Johnson.
\newblock Spiral: A generator for platform-adapted libraries of signal
  processing alogorithms.
\newblock {\em The International Journal of High Performance Computing
  Applications}, 18(1):21--45, 2004.

\bibitem{linear}
{R Core Team}.
\newblock {\em R: A Language and Environment for Statistical Computing}.
\newblock R Foundation for Statistical Computing, Vienna, Austria, 2018.

\bibitem{rasch2017atf}
A.~Rasch~et al.
\newblock Atf: A generic auto-tuning framework.
\newblock In {\em High Performance Computing and Communications}. IEEE, 2017.

\bibitem{rinard2006probabilistic}
M.~Rinard.
\newblock Probabilistic accuracy bounds for fault-tolerant computations that
  discard tasks.
\newblock In {\em Proceedings of the 20th annual international conference on
  Supercomputing}, pages 324--334. {ACM}, 2006.

\bibitem{Roustant2012}
O.~Roustant~et al.
\newblock {DiceKriging, DiceOptim: Two R Packages for the Analysis of Computer
  Experiments by Kriging-Based Metamodeling and Optimization}.
\newblock {\em Journal of Statistical Software}, 2012.

\bibitem{kriging}
J.~Sacks, W.~J. Welch, T.~J. Mitchell, and H.~P. Wynn.
\newblock Design and analysis of computer experiments.
\newblock {\em Statistical Science}, 4(4):409--423, 1989.

\bibitem{samadi2014paraprox}
M.~Samadi, D.~A. Jamshidi, J.~Lee, and S.~Mahlke.
\newblock Paraprox: Pattern-based approximation for data parallel applications.
\newblock {\em ACM SIGPLAN Notices}, 49(4):35--50, 2014.

\bibitem{samadi2013sage}
M.~Samadi~et al.
\newblock Sage: Self-tuning approximation for graphics engines.
\newblock In {\em MICRO}. IEEE, 2013.

\bibitem{schaefer2009atune}
C.~A. Schaefer, V.~Pankratius, and W.~F. Tichy.
\newblock Atune-il: An instrumentation language for auto-tuning parallel
  applications.
\newblock In {\em European Conference on Parallel Processing}, pages 9--20.
  Springer, 2009.

\bibitem{Shen:2013:GFA:2482767.2482785}
J.~Shen, A.~L. Varbanescu, H.~Sips, M.~Arntzen, and D.~G. Simons.
\newblock Glinda: A framework for accelerating imbalanced applications on
  heterogeneous platforms.
\newblock In {\em Proceedings of the ACM International Conference on Computing
  Frontiers}, CF '13, pages 14:1--14:10, New York, NY, USA, 2013. ACM.

\bibitem{Stone1997}
C.~J. Stone~et al.
\newblock {Polynomial splines and their tensor products in extended
  linearmodeling}.
\newblock {\em Annals of Statistics}, 1997.

\bibitem{sui2016proactive}
X.~Sui~et al.
\newblock Proactive control of approximate programs.
\newblock {\em ACM SIGOPS Operating Systems Review}, 2016.

\bibitem{vuduc2005oski}
R.~Vuduc, J.~W. Demmel, and K.~A. Yelick.
\newblock Oski: A library of automatically tuned sparse matrix kernels.
\newblock In {\em Journal of Physics: Conference Series}, volume~16, page 521.
  IOP Publishing, 2005.

\bibitem{WEBSTER1980}
R.~Webster~et al.
\newblock Optimal interpolation and isarithmic mapping of soil properties {III}
  changing drift and universal kriging.
\newblock {\em Journal of Soil Science}, 1980.

\bibitem{whaley1998automatically}
R.~C. Whaley and J.~J. Dongarra.
\newblock Automatically tuned linear algebra software.
\newblock In {\em Proceedings of the 1998 ACM/IEEE conference on
  Supercomputing}, pages 1--27. IEEE Computer Society, 1998.

\bibitem{tidyverse}
H.~Wickham.
\newblock {\em tidyverse: Easily Install and Load the 'Tidyverse'}, 2017.
\newblock R package version 1.2.1.

\end{thebibliography}

\end{document}